\documentclass[twocolumn,letter,superscriptaddress]{revtex4-1}
 \newcommand{\bra}[1]{\left< #1\right|}
\newcommand{\ket}[1]{\left|#1 \right>} 
\newcommand{\EqLabel}[1]{\label{#1}}

\newcommand{\Mona}{}

\usepackage{hyperref}
\usepackage{graphicx}
\usepackage{amsmath}
\usepackage{xcolor}
\usepackage{bm}
\usepackage{color}
\usepackage{enumitem} 
\usepackage[T1]{fontenc}
\usepackage{lipsum}
\usepackage{amsmath, array} 
\usepackage[normalem]{ulem}
\usepackage{soul,xcolor}
\usepackage{lineno}

\definecolor{darkblue}{rgb}{0.0,0.0,0.0}
\definecolor{darkblue2}{rgb}{0.0,0.0,0.0}

\begin{document}
\title{Light Bipolarons Stabilized by Peierls Electron-Phonon Coupling} 

\author{John Sous}\thanks{jsous@phas.ubc.ca}  \affiliation{\!Department \!of \!Physics and
  Astronomy, \!University of\!  British Columbia, \!Vancouver, British
  \!Columbia,\! Canada,\! V6T \!1Z1} \affiliation{\!Department \!of Chemistry, \!University
  of\!  British Columbia, \!Vancouver, British \!Columbia,\! Canada,\!
  V6T \!1Z1} 
   \affiliation{\!Stewart Blusson Quantum Matter \!Institute, \!University
  of British Columbia, \!Vancouver, British \!Columbia, \!Canada,
  \!V6T \!1Z4}
\affiliation{\!ITAMP, \!Harvard-Smithsonian \!Center \!for \!Astrophysics, \! \!Cambridge, \!Massachusetts, \!02138, \!USA}
\affiliation{\!Department \!of \!Physics, \!Harvard\!  University, \!Cambridge, \!Massachusetts, \!02138, \!USA}
\author{Monodeep Chakraborty} \affiliation{Centre \!for \!Theoretical \!Studies, \!Indian \!Institute \!of \!Technology, \!Kharagpur, \!India}
\author{Roman V. Krems}
\affiliation{\!Department \!of Chemistry, \!University
  of\!  British Columbia, \!Vancouver, British \!Columbia,\! Canada,\!
  V6T \!1Z1} 
\author{Mona Berciu}\thanks{berciu@phas.ubc.ca}
\affiliation{\!Department \!of \!Physics and Astronomy, \!University
  of\!  British Columbia, \!Vancouver, British \!Columbia,\! Canada,\!
  V6T \!1Z1} \affiliation{\!Stewart Blusson Quantum Matter \!Institute, \!University
  of British Columbia, \!Vancouver, British \!Columbia, \!Canada,
  \!V6T \!1Z4}

\begin{abstract}
  It is widely accepted that phonon-mediated high-temperature
  superconductivity is impossible at ambient pressure, because of the
  very large effective masses of polarons/bipolarons at strong
  electron-phonon coupling. Here we challenge this belief by showing
  that strongly bound yet very light bipolarons appear for strong
  Peierls/Su-Schrieffer-Heeger coupling. These
  bipolarons also exhibit many other unconventional properties, {\em e.g.} at strong coupling there are two low-energy bipolaron bands
  that are stable against strong Coulomb repulsion. Using numerical simulations and analytical arguments,
  we show that these
  properties result from the specific form of the phonon-mediated
  interaction, which is of ``pair-hopping'' instead of regular
    density-density type. {\Mona This unusual effective interaction is
    bound to have non-trivial consequences for the superconducting
    state expected to arise at finite carrier concentrations, and
    should favor a large critical temperature.}
\end{abstract}

\maketitle

{\em Introduction}.---Since the discovery of superconductivity in Hg
with critical temperature $T_c=4.2$ K \cite{Firstsuperconductor}, the quest for materials
with high $T_c$ has been a central driver of research in condensed
matter physics, leading to the discovery of many other
superconductors including the unconventional ``high''-$T_c$ cuprate
\cite{cuprate_super} and iron-based
\cite{Ironbased_super1,Ironbased_super2} families, besides many
conventional and unconventional low-$T_c$ ones.
  
 Conventional low-$T_c$ superconductivity is understood to be a
 consequence of electron-phonon coupling \cite{BCS1, BCS2}: Exchange
 of phonons binds electrons into Cooper pairs \cite{CooperPair} which
 condense into a superfluid. While there is no proven theory of
 high-$T_c$ superconductivity, it is widely accepted that
 phonon-mediated superconductivity cannot exhibit high $T_c$ (at
 ambient pressure).  High $T_c$ would require strong electron-phonon
 coupling, but in this limit the electrons become dressed by clouds of
 phonons forming polarons, with a renormalized effective mass
 \cite{Landau, Landau_Pekar, Froh1, Holstein1, Froh2, Holstein2,
   Feynman1,
   Feynman2,Prokofiev_FrohPolaron1,Prokofiev_FrohPolaron2,Kornilovitch_polaron_general1,Bonca_polaron1,Bonca_polaron2,Kornilovitch_polaron_dimensionality2,
   Holstein_MA}. As the coupling strength increases, the effective
 mass grows faster than the phonon-mediated binding, resulting in 
 suppression of $T_c$ \cite{ImpossibleBipolaronicSuperconductivity}.
 In other words, it is generally believed impossible to form bipolarons
(polaron pairs bound by phonon exchange) that remain light at
 strong electron-phonon coupling, making high-$T_c$ bipolaronic superconductivity
 very unlikely \cite{ImpossibleBipolaronicSuperconductivity}, {\color{darkblue2}{\footnote{High-$T_c$ bipolaronic superconductivity may be possible in very special situations discussed in Ref. \cite{HagueSuperLight}}. }}

Such arguments, however,
are based on studies of the Holstein \cite{Holstein2} and Fr\"ohlich
\cite{Froh1, Froh2} models.  There, phonons modulate the
potential energy of the electrons, which explains why polarons and
bipolarons become heavier as the coupling strength increases.  On the other hand, the
coupling to phonons may also modulate the hopping integrals, as
described by the  Peierls model \cite{SSH1a,SSH1b,SSH1c} (known as
the Su-Schrieffer-Heeger (SSH) model for polyacetylene
\cite{SSH2a,SSH2b}).  Recently, it was shown that single polarons in
this latter class of models can be light at strong coupling strengths
\cite{Dominic}.

Here we study for the first time phonon-mediated binding of electrons
into bipolarons in the Peierls model. We show that Peierls
electron-phonon coupling leads to strong phonon-mediated attraction
between electrons, which results in the formation of strongly
bound yet very light bipolarons: Their mass 
at strong coupling is close to twice the free electron mass.  Such light
bipolarons are expected to condense into a superfluid  at very
high temperatures \cite{ImpossibleBipolaronicSuperconductivity,HagueSuperLight}. Our work thus points to a new direction in the
search for high-$T_c$ superconductors: Designing materials with
electron-phonon coupling predominantly of the Peierls-type can lead to
phonon-mediated superconductivity at high temperatures.

{\em Model and methods}.---We study the singlet state of two
spin-${1\over 2}$ fermions in an infinite one-dimensional chain
described by the Hamiltonian ${\cal H}={\cal H}_{\rm e} + {\cal
  H}_{\rm ph} + \hat{V}_{\rm{e-ph}}$, where:
\begin{eqnarray}
{\cal H}_{\rm e} = -t \sum_{i,\sigma}^{}\left( c_{i,\sigma}^\dagger
c_{i+1,\sigma} + h.c.\right) +  \sum_{i}^{} U(\delta) \hat{n}_{i,\uparrow}
\hat{n}_{i+\delta,\downarrow} \quad
\end{eqnarray}
is the {\color{darkblue2}{extended Hubbard model of bare electrons with on-site $U(0)=U$ and nearest-neighbor $U(1)=V$ screened repulsion, }}
$i$ is the site index and $\hat n_{i,\sigma} = c_{i,\sigma}^\dagger
c_{i,\sigma}$ counts particles with spin $\sigma$ at site $i$.  ${\cal
  H}_{\rm ph} = \Omega \sum_{i}^{} b_{i}^\dagger b_i$ (in units of
$\hbar$) is the phonon Hamiltonian describing a single Einstein mode
with frequency $\Omega$, and
\begin{equation}
\EqLabel{4} \hat{V}_{\rm{e-ph}}=g\sum_{i,\sigma}^{}\left(
c_{i,\sigma}^\dagger c_{i+1,\sigma} + h.c.\right)\left(
b_i^\dagger+b_i - b_{i+1}^\dagger-b_{i+1}\right) 
\end{equation}
is the Peierls/SSH electron-phonon coupling \cite{Dominic}. We
characterize the electron-phonon strength using the dimensionless
effective coupling $\lambda = 2g^2/(\Omega t)$.  We investigate the
singlet eigenstates using variational exact diagonalization (VED) \cite{Bonca_bipolaron1, Bonca_bipolaron2, Monodeep_bipolaron}
and an extension of the Momentum Average (MA) approach \cite{Mona_MA, Holstein_MA, Dominic, Repulsive, SousLongPaper}.

\begin{figure}[t]
\includegraphics[width=\columnwidth]{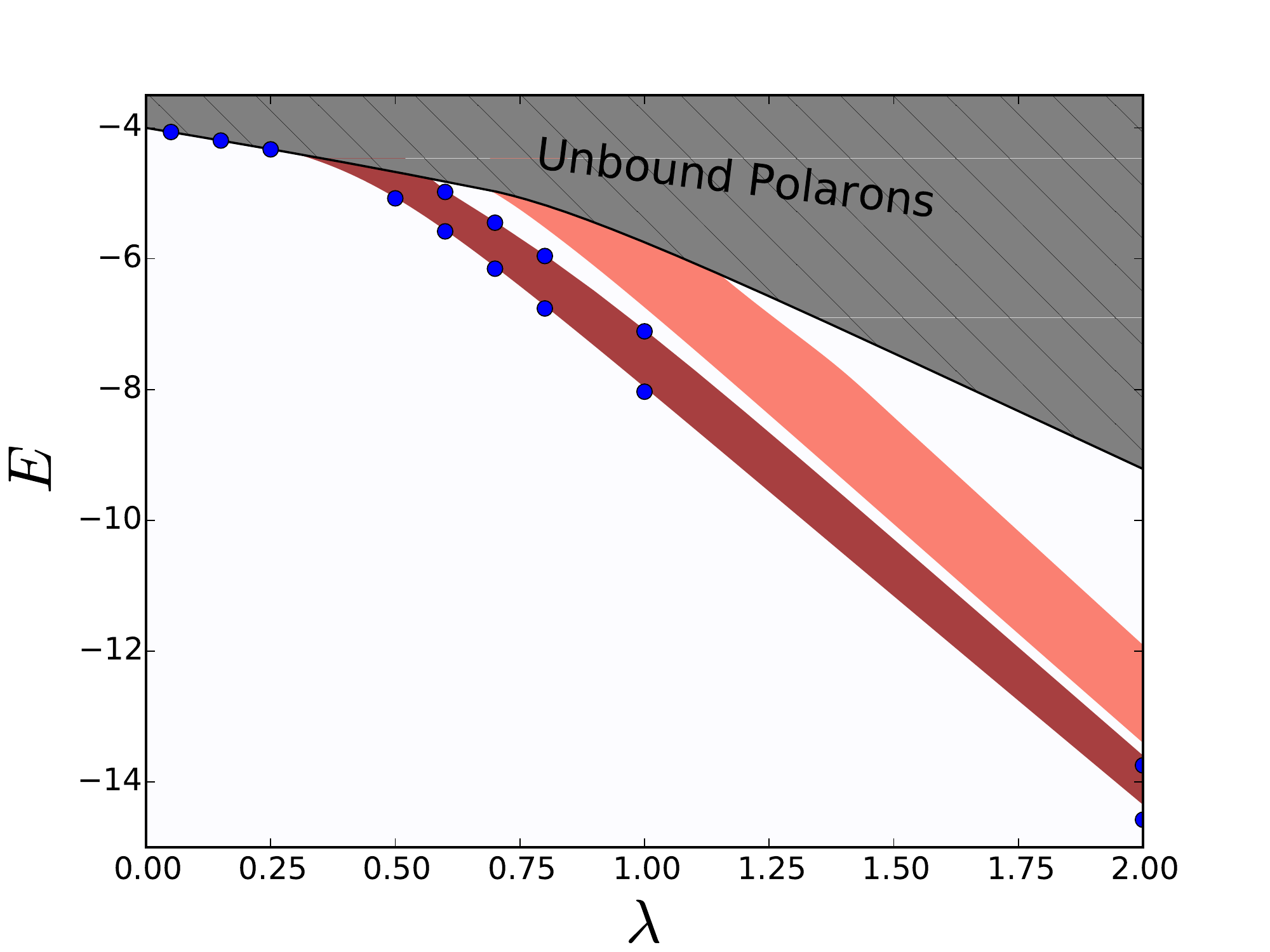}
\caption{{(Color online) Two-polaron phase diagram for $U(\delta)=0$ and $\Omega=3$.} The
  diagram represents the evolution of the low-energy region of the
  singlet sector with $\lambda$. Energies are in units of $t$. The
  dimensionless effective coupling is $\lambda = 2g^2/(\Omega t)$. The
  shaded grey area shows the lower part of the two-polaron continuum.
  The dark red region represents the lowest energy bipolaron band,
  while the salmon region represents the higher energy bipolaron band.
  These results were obtained with MA and are in good agreement with
  VED results (blue circles) shown for the low-energy bipolaron.}
\label{fig1}
\end{figure}


{\em Numerical results}.---We first set $U(\delta)=0$ and investigate the
stability and properties of the resulting bipolarons. The role played
by $U(\delta)$ is discussed later.

Figure \ref{fig1} shows the evolution with $\lambda$ of the low-energy
region of the singlet sector, for $U(\delta)=0$ and $\Omega=3$ (all energies are in
units of $t=1$). The shaded grey area shows the lower part of the
two-polaron continuum: these states describe two unbound polarons,
their energies being the convolution of two single polaron spectra.
The dark red region shows the location of the lowest bipolaron band.
VED confirms its existence for all $\lambda > 0$, although for weak coupling
$\lambda \lesssim 0.3$, the bipolaron ground state lies just below the continuum
and cannot be resolved on this scale. With increasing $\lambda$, the
bipolaron band moves further below the continuum and for $\lambda \gtrsim 0.57$ it
becomes fully separated from it. For strong coupling $\lambda > 1$, this
band is accompanied by a higher energy bipolaron band (salmon-colored
region), whose evolution with $\lambda$ closely mirrors that of the lower
band, suggesting a common origin. Note that this
second band lies below the bipolaron$+$one-phonon continuum (not
shown) that starts at $\Omega$ above the ground state, and therefore it is
an infinitely lived bipolaron.

\begin{figure}[t]
\includegraphics[width=\columnwidth]{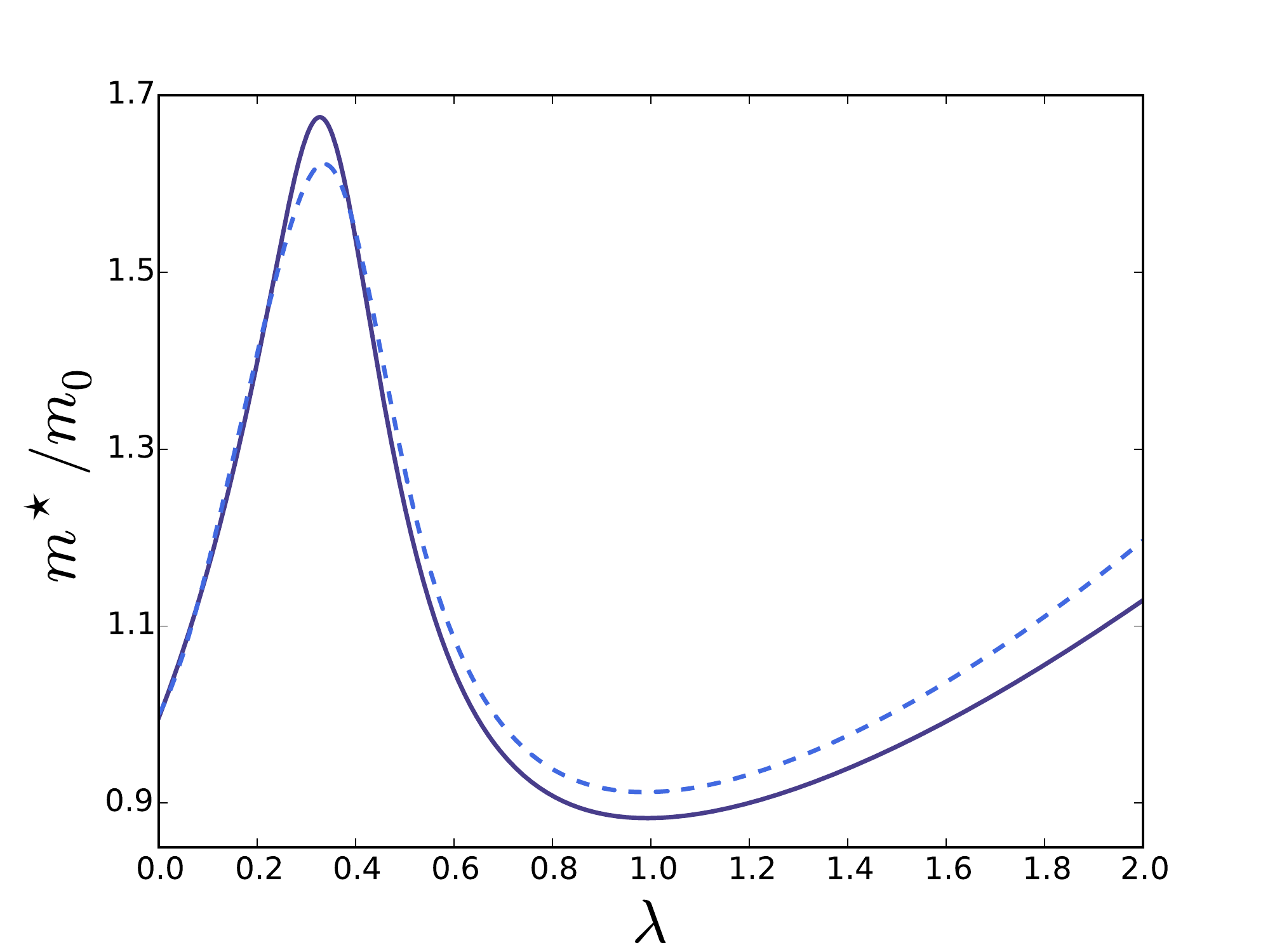}
\caption{{(Color online) Dependence of the effective mass of the
    low-energy bipolaron on $\lambda$, for $U(\delta)=0$ and $\Omega = 3.0$.} $m_0= 2m_e$
  is twice the free electron mass. The bipolaron's effective mass is
  defined as $m^* = \Big(\frac{\partial ^2 E_{BP}(K)}{\partial K^2} \Big)^{-1}
  \Big|_{K = K_{GS}}$. The solid (dashed) lines are VED (MA) results.
  Note that $m^* \sim 2 m_e$ in the strongly coupled regime, $\lambda > 1$. }
\label{fig2}
\end{figure}
 
Clearly the bandwidths of both bipolaron bands are wide even at
extremely strong couplings $\lambda \sim 2$ (this persists for $\lambda > 2$ but such
values are unphysical), showing that the bipolarons remain light even
when very strongly bound. This is further confirmed in Figure
\ref{fig2}, where we plot the low-energy bipolaron's effective mass
$m^*$, in units of two free particle masses, $m_0=2m_e=\hbar^2/ta^2$,
where $a$ is the lattice constant.  \textcolor{darkblue}{$m^*$ varies non-monotonically with $\lambda$, with a peak at $\lambda \sim 0.325$ where the bipolaron ground state energy starts to drop fast below the lower edge of the two-polaron continuum (see Figure \ref{fig1}), {\em i.e.} the bipolaron crosses over into the strongly bound regime. }Importantly, the ratio $m^*/m_0$ stays close to 1 for $\lambda \gtrsim 1$. In other words, the Peierls
bipolaron's effective mass remains comparable to that of a pair of
free fermions even at very strong coupling $\lambda = 2$. For comparison,
for the same $\Omega$ and $\lambda=2$, the Holstein bipolaron's bandwidth is
$0.0135$, {\em i.e.} its mass is larger by about two orders of
magnitude.

The existence of strongly bound yet light Peierls bipolarons at strong
coupling is our central result.

We now discuss the second bipolaron band. For reference, we note that
the one-dimensional Holstein and Fr\"ohlich models host only one
bipolaron band within $\Omega$ of the ground-state energy (if $U=0$)
\cite{Bonca_bipolaron2}. This is precisely what is generically
expected. For Holstein coupling, the phonon-mediated effective
interaction can be modeled as an effective on-site attraction $-
\Delta E \sum_{i}^{}\hat{n}_{i\uparrow} \hat{n}_{i\downarrow}$, where
$\Delta E \rightarrow 2g^2/\Omega$ as $\lambda \rightarrow \infty$
\cite{Bonca_bipolaron1, Bonca_bipolaron2}. In one dimension, such
effective on-site attraction  binds two fermions into a singlet
state, but there is only one bound state. The existence of a second bipolaron
state is thus very surprising, and points to a new mechanism
behind pairing.

 \begin{figure}[t]
\includegraphics[width=\columnwidth]{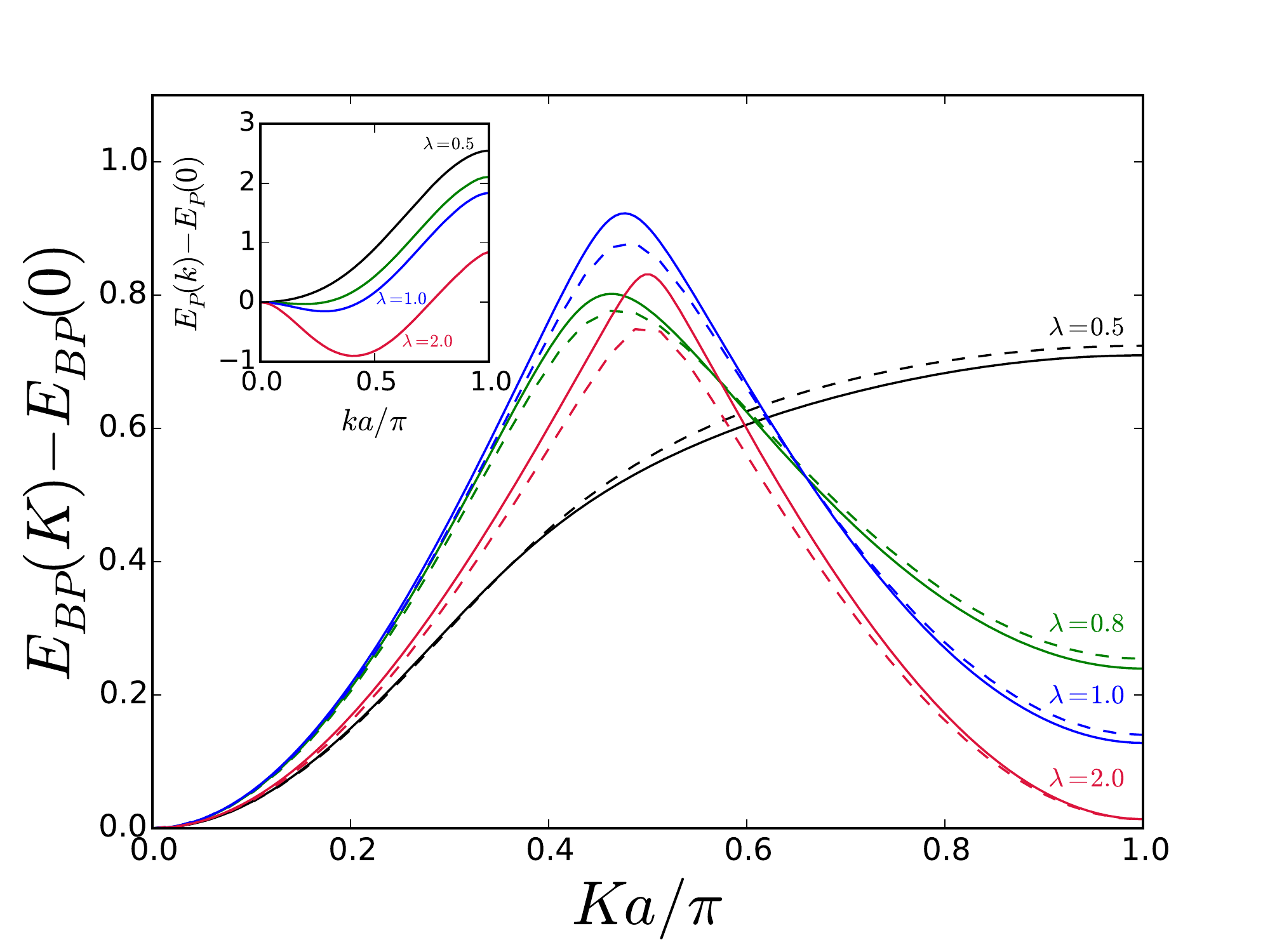}
\caption{{(Color online) Dispersion $E_{BP}(K) - E_{BP}(0)$ of the low-energy
    bipolaron, for various values of $\lambda = 2g^2/(\Omega t)$ at
    $U(\delta)=0$ and $\Omega = 3$. The inset shows the polaron dispersion
    $E_{P}(k) - E_{P}(0)$ for the same parameters.} All energies are
  in units of $t$. In the main figure, solid lines are VED results and
  dashed lines are MA results. Results in the inset were obtained with
  MA, and are in good agreement with numerical results \cite{Dominic}.
}
\label{fig3}
 \end{figure}
 
To understand this new physics, we first consider the dispersion of
the lowest energy bipolaron, and its evolution with increasing
$\lambda$. This is shown in Figure \ref{fig3} for the positive half of
the Brillouin zone. The curves have been shifted for ease of
comparison (their absolute positions can be inferred from Figure
\ref{fig1}). The inset shows the polaron dispersions for the same
coupling parameters.

\begin{figure}[t]
\includegraphics[width=\columnwidth]{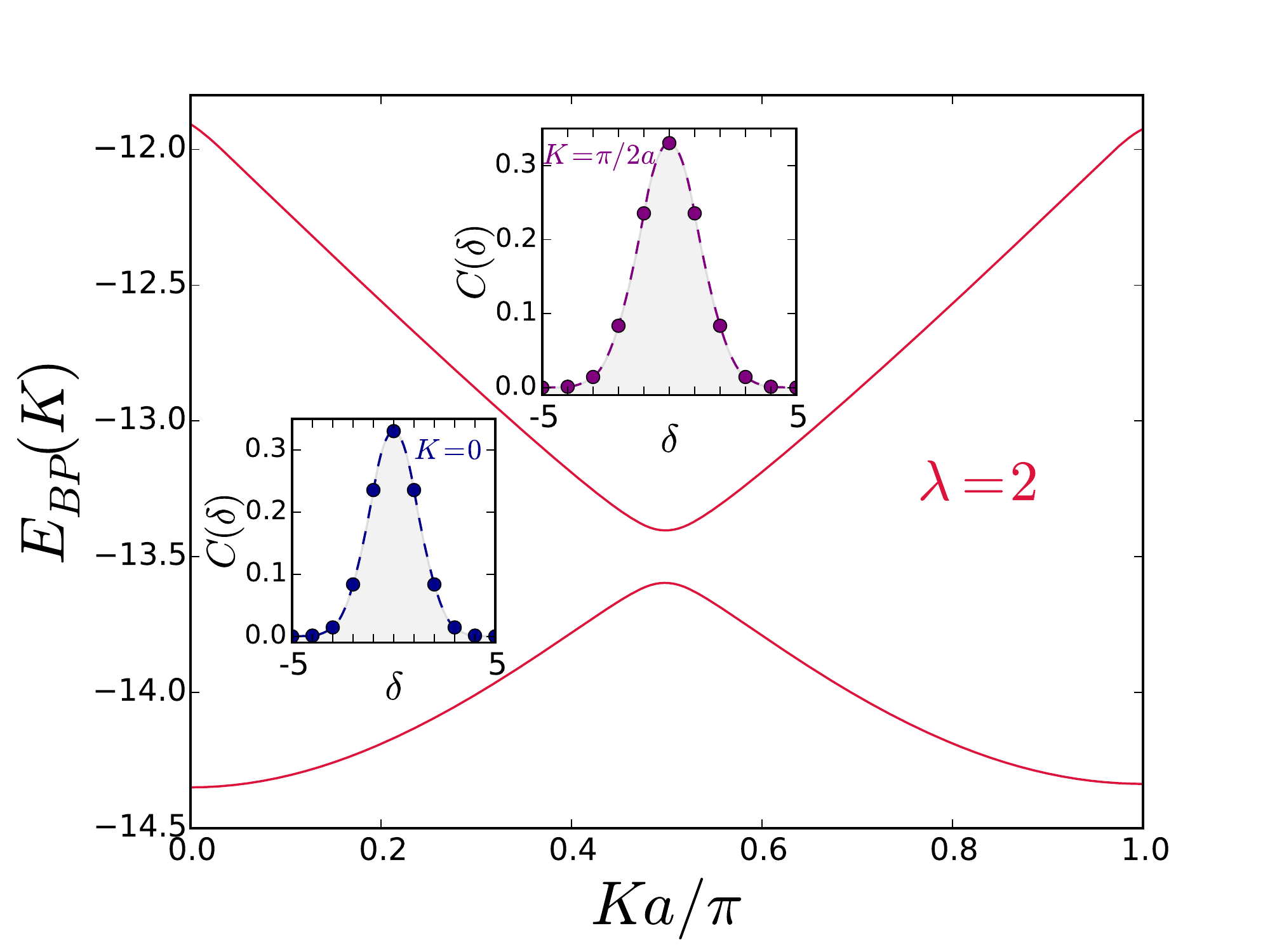}
\caption{{(Color online) Dispersion $E_{BP}(K)$ of both bipolaron bands, for
    $U(\delta)=0$, $\Omega = 3$ and $\lambda = 2$, showing an avoided
    crossing.} $E_{BP}(K)$ is in units of $t$. These are MA results. \color{darkblue2}{The insets show the 
    low-energy bipolaron spatial correlation function $C(\delta)=\bra{\Psi_{BP}}\frac{1}{N}\sum_i \hat{n}_{i,\uparrow}\hat{n}_{i+\delta,\downarrow}\ket{\Psi_{BP}}$} at $\lambda = 2$ for $K=0$ and $K=\pi/2a$ obtained with VED. In both cases, 
    the electrons in the bipolaron wavefunction are found with large probability to be up to two sites apart.}
\label{fig4}
\end{figure}

At couplings $\lambda \leq 0.5$ where only this bipolaron band exists,
the dispersion has the standard behavior, being monotonically increasing with
$K$. For larger $\lambda$,  the dispersion has a
rather unusual shape, strongly peaked near $Ka = {\pi\over 2}$. This
shape is highly suggestive of an avoided crossing with a band located
above (the second bipolaron state that emerges at these couplings).
This is confirmed when we plot both bands for $\lambda=2$ in Figure
\ref{fig4}. The gap that opens between the two bands
varies only weakly with $\lambda$, see Figure \ref{fig1}. This
behavior suggests the existence of two bound states
with different symmetries, coupled by a $\lambda$-independent
symmetry-breaking term.

{\em Analytical arguments}.---To unravel the pairing mechanism and explain the origin of the two
bipolaron states and their avoided crossing, we consider the
anti-adiabatic limit $\Omega \gg t,g$. We obtain analytical results
by projecting out the high-energy Hilbert subspaces with one or more
phonons \cite{Projection_technique}. Note that strong-coupling $\lambda
\gg 1$ is included within the anti-adiabatic regime if $t\ll g \ll
\Omega$ such that $g^2 \gg \Omega t$.

As discussed in Ref. \cite{Dominic}, the effective Hamiltonian in the
single particle sector is
\begin{eqnarray}
\hat{h}_1 = -\epsilon_0\sum_{i,\sigma}^{}\hat{n}_{i,\sigma} -
\sum_{i,\sigma}^{} \left( t c_{i,\sigma}^\dagger c_{i+1,\sigma}- t_2
c_{i,\sigma}^\dagger c_{i+2,\sigma} + h.c.\right) \nonumber
\end{eqnarray}
In addition to the nearest-neighbor bare particle hopping, $\hat{h}_1$
contains the polaron formation energy $\epsilon_0 = 4g^2/\Omega$, and
a dynamically generated next-nearest-neighbor hopping $t_2=g^2/\Omega$
resulting from virtual emission and subsequent absorption of a phonon
by the particle, as it hops on and off an intermediate site. This term
becomes dominant for large $\lambda$ and explains the change in the
shape of the polaron dispersion $E_P(k) = -\epsilon_0 -2t \cos(ka) +
2t_2\cos(2ka)$ observed in the inset of Figure \ref{fig3} (for
detailed discussions see Ref. \cite{Dominic}).

In the singlet sector, we find that the effective two-particle
Hamiltonian is: $\hat{h}_{2,s} =\hat{h}_1 + \hat{U}_{0,2} +
\hat{U}_{1}$. The additional terms describe short-range  phonon-mediated 
interactions between the polarons. Specifically
\begin{align}
\hat{U}_{0,2}  = -T_{0,0}\sum_{i}^{}\left[
    c^\dagger_{i-1,\uparrow}c^\dagger_{i-1,\downarrow}c_{i,\downarrow}
    c_{i,\uparrow}+ h.c.\right] \hspace{10mm} \nonumber \\  + T_{0,2} \sum_{i}^{}
  \left[\left( c^\dagger_{i+1,\uparrow}c^\dagger_{i-1,\downarrow}-
    c^\dagger_{i+1,\downarrow}c^\dagger_{i-1,\uparrow}\right)c_{i,\downarrow}
    c_{i,\uparrow} + h.c. \right] \nonumber
\end{align}
describes nearest-neighbor ``pair-hopping'' of an on-site singlet with
$T_{0,0}={4g^2\over \Omega}$, and transitions between on-site and
next-nearest-neighbor singlets with $T_{0,2} = {2g^2\over \Omega}$.
They arise through emission and absorption of a  phonon,
{\em e.g.} $c^\dagger_{i,\uparrow}c^\dagger_{i,\downarrow}|0\rangle
\stackrel{\hat{V}_{\rm{e-ph}}}{\Longrightarrow}c^\dagger_{i+
  1,\uparrow}c^\dagger_{i,\downarrow} b^\dagger_{i+1}|0\rangle
\stackrel{\hat{V}_{\rm{e-ph}}}{\Longrightarrow} c^\dagger_{i+1,\uparrow}
c^\dagger_{i+1,\downarrow}|0\rangle$ allows one particle to hop by
emitting a phonon, then the second particle absorbs the phonon
and hops to its partner's new site. 
This is one of the processes contributing to $T_{0,0}$; all relevant
processes can be similarly inferred.

The other effective interaction term
\begin{eqnarray*}
& \hat{U}_{1} & = +T_{1,1}\sum_{i,\sigma}^{}\left[
    c^\dagger_{i+1,\sigma}c^\dagger_{i+2,-\sigma}c_{i+1,-\sigma}
    c_{i,\sigma}+ h.c.\right] \\ && +J \sum_{i,\sigma}^{}
  c^\dagger_{i+1,\sigma}c^\dagger_{i,-\sigma}c_{i,\sigma}c_{i+1,-\sigma}
\end{eqnarray*}
acts when the particles are on adjacent sites and describes the
pair-hopping of a nearest-neighbor singlet with $T_{1,1} = {2g^2\over
  \Omega}$, and an antiferromagnetic $xy$ exchange  with $J={4g^2\over \Omega}$.

Note that none of these terms are of the density-density type of
interaction that is assumed to be the functional form for
phonon-mediated effective interactions. More specifically, these terms
can be written in the form $\sum_{k,k',q} u(k+k',q) c^\dagger_{k+q,\uparrow}
c^\dagger_{k'-q,\downarrow} c_{k',\downarrow} c_{k,\uparrow}$ allowed by translational invariance.
The interaction vertex, $u(k+k',q)$, depends not only on the exchanged
momentum, $q$, as is usually assumed to be the case, but also on the
total momentum of the interacting pair, $k+k'$. It is therefore important
to understand the consequences of such interactions, for example,
how they affect the properties of BCS- or
Bose-Einstein-Condensate(BEC)-type superconductors. We leave such
studies for future work.

The origin of the two different symmetry states leading to the two
bipolaron bands is now clear. First, let us set $t=0$. In this case,
the low-energy Hilbert subspace factorizes into two sectors, with the particles
being separated either by an even or by an odd number of sites; the
remaining terms in the Hamiltonian do not mix these subspaces. To 
solve for bound states, we calculate the two-particle propagator \cite{MBmp} 
and check for discrete poles appearing below the continuum. We find 
that $\hat{U}_{0,2}$ and  $\hat{U}_{1}$ can lead to the appearance of a bound 
state in their respective subspace. The former has a monotonically 
increasing dispersion, $E_{\rm even}(K) = -2\epsilon_0 - 2T_{0,0}\cos(Ka) + \mu(K)$, 
where $\mu(K) = \frac{F_{\rm even}(K)}{2 \theta(K)} - \frac{1}{2} \sqrt{(\frac{F_{\rm even}(K)}{\theta(K)})^2 + 4 \zeta(K)}$, 
with $F_{\rm even}(K) = 2T_{0,0}\cos(Ka)$, $\theta(K) = 1 - \frac{(f_2(K))^2}{\alpha(K)}$, $\zeta(K) = \frac{\alpha(K)}{\theta(K)}$
; $f_2(K) = 2t_2 \cos(Ka)$ and $\alpha(K) = 2 (T_{0,2}+f_2(K))^2$. The 
latter has a monotonically decreasing dispersion, $E_{\rm odd}(K) =
-2\epsilon_0 - J + 2(t_2 +T_{1,1})\cos(Ka) + \kappa(K)$, with $\kappa(K) =
\frac{(f_2(K))^2}{f_2(K) + F_{\rm odd}(K)}$, where $F_{\rm odd}(K) =
-J + 2T_{1,1}\cos(Ka)$. Note that both
these energies are controlled by the energy scale $g^2/\Omega$,
explaining why they evolve similarly with increasing $\lambda$. When
$t$ is turned on, the nearest-neighbor hopping term breaks this symmetry
and leads to the avoided crossing, and hence the two bipolaron bands
with unusual dispersions shown in Figure \ref{fig4}.

\begin{figure}[t]
\includegraphics[width=88mm]{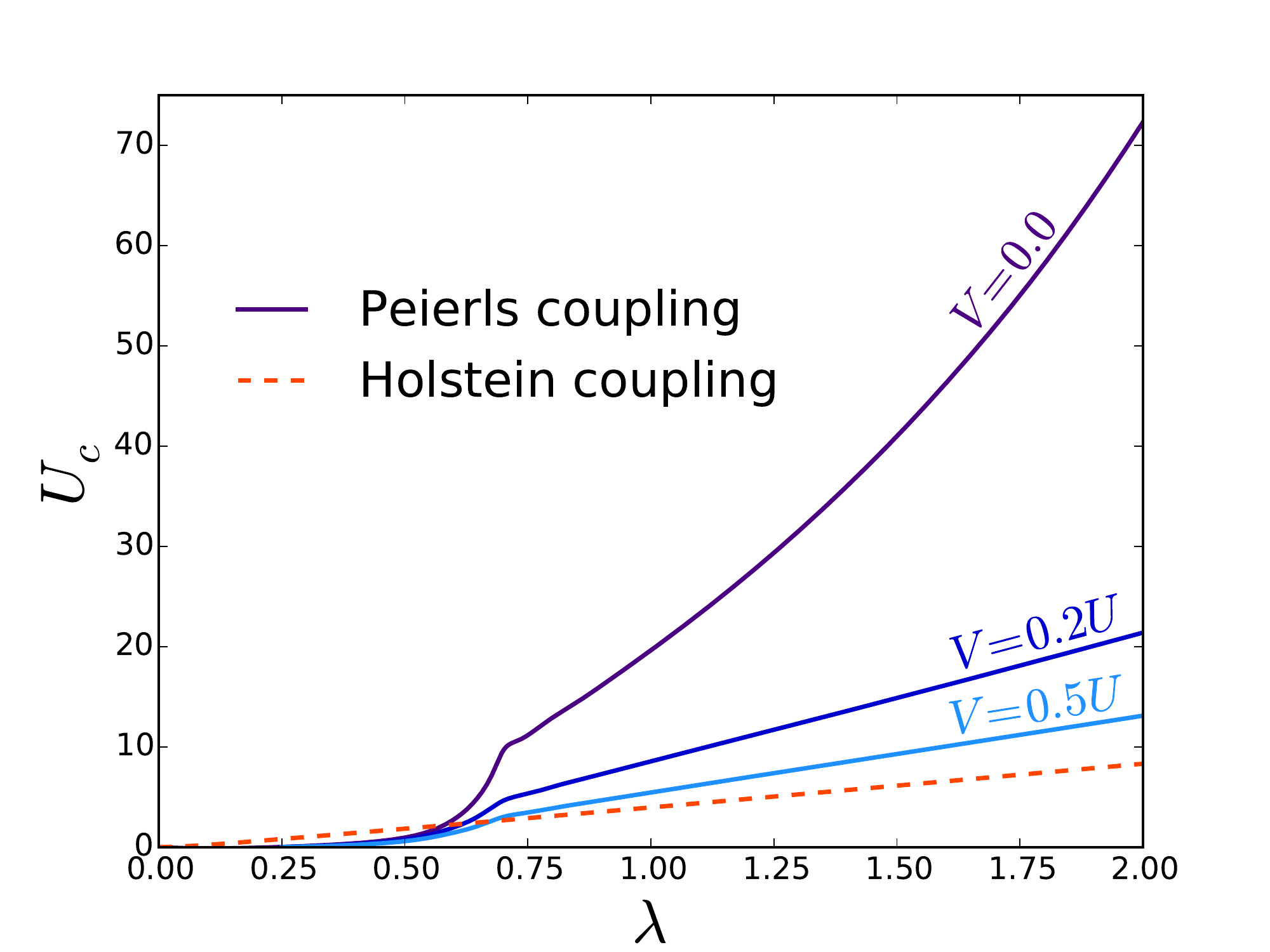}
\caption{{(Color online) $U_c$-$\lambda$ stability diagram for the Peierls/SSH
    (solid lines) and Holstein (dashed line)
    bipolarons at $\Omega = 3$.} $U_c$ is in units of $t$. For the Peierls/SSH coupling, $\lambda = 2g^2/(\Omega
  t)$, while for the Holstein coupling, $\lambda = g_{\rm
    H}^2/(2\Omega t)$, where $g_{\rm H}$ is the Holstein
  electron-phonon coupling. 
   These are VED results, and reveal a qualitative difference between the
  stability of the two types of bipolarons at strong coupling
   $\lambda > 1$. {\color{darkblue2}{The Peierls bipolaron remains more stable than the Holstein bipolaron even in the presence of a screened nearest-neighbor Coulomb repulsion $V=0.2U$ and $V=0.5U$.}}
}
\label{fig5}
\end{figure}


We now address the role of the Coulomb repulsion $U(\delta)$. In
Figure \ref{fig5}, we display the critical value $U_c$ above which
bipolarons dissociate into unbound polarons, for the Peierls/SSH
(solid lines) and Holstein (dashed line) models. Clearly, $U_c$ is much
larger for Peierls bipolarons than for Holstein bipolarons, {\color{darkblue2}{even with a strong nearest-neighbor $V=0.5U$.}} This
is yet another qualitative difference between the two models.

In the Holstein model, $U$ directly competes with the on-site
attraction $\Delta E$ mediated by phonons. A smooth crossover from an
on-site bipolaron to a weakly bound bipolaron with the particles
on neighboring sites is observed for $U\sim \Delta E$, and a somewhat larger
$U$ suffices to dissociate the bipolaron \cite{Bonca_bipolaron1}.

For the Peierls/SSH coupling, consider again the anti-adiabatic limit
with $t=0$. Here, for $V=0$, a sufficiently large $U$ will destabilize
the bound state in the even sector, but will have much less effect on the
bound state of the odd sector. Hybridization due to a finite $t$ will
then result in a low-energy bipolaron similar to the bound state
of the odd sector. Consequently, one expects a stable bipolaron
even for large values of $U$. Moreover, for a sufficiently large
$U$ value, one expects a transition to a bipolaron with ground state momentum
$K_{GS} = \pi/a$, favored by the odd bound state. Indeed, we verified this behavior
 in the anti-adiabatic limit (not shown). \textcolor{darkblue2}{Including a nearest-neighbor 
 repulsion $V \sim U$ further suppresses $U_c$, as verified in Figure \ref{fig5}.}

Away from the anti-adiabatic limit, see Figure \ref{fig5}, we find that
the $t$-controlled mixing between even and odd bound states suffices
to destabilize both states, at large enough $U>U_c$.
\textcolor{darkblue2}{Still, $U_c$ is larger than for the Holstein model even for a very strong $V/U=0.5$.}

\textcolor{darkblue}{The Peiers bipolarons are thus stable
 in a much wider range of repulsive $U$ than the Holstein bipolarons.}
This is a direct consequence of the existence of the two
bound states with different symmetries, one of which is 
only weakly affected by large $U$ (at $t=0$).  


{\em Summary and discussion}.---We have demonstrated the existence of
strongly bound yet light Peierls bipolarons, {\color{darkblue2}{stable against
large values of the screened Coulomb repulsion.
The light bipolaron is a consequence of the Peierls-type coupling, not of 
special circumstances like in Ref. \cite{HagueSuperLight}, making our conclusions applicable to
a large class of systems.}}
We explained that pairing is mediated by pair-hopping
terms instead of the customary attractive Hubbard-like terms. This
unusual attraction binds two low-energy bipolaron states, instead of
one. As a result of an avoided crossing, these Peierls bipolarons have
unique dispersions.

The binding mechanism
poses questions about the nature of superconductivity at finite
carrier densities in higher dimensions: Light bipolarons should
condense into a BEC-type superconductor with high $T_c$. 
This should be relevant to conjugated polymers \cite{SSH2a,SSH2b,SSH_Polymers}, organic semiconductors
\cite{Organic_SSH, Triosi1, Candia, Triosi2}, some oxides \cite{SemiconductorOxides_review, Steve_Berciu_Sawatzky},
and engineered quantum simulators \cite{polar-molecules1,polar-molecules2,rydbergs3,d-wave1, d-wave4,Lobo_SSHfermionicsimulator}.
Recent work claims a record $T_c$ for superconductivity
in doped organic p-terphenyl molecules
\cite{RecordHighTcsuperconductivity1,RecordHighTcsuperconductivity2},
where the Peierls coupling is important, and attributes it to a possible
bipolaronic mechanism \cite{RecordHighTcsuperconductivity1}. 
\textcolor{darkblue}{Similarly, our work may be relevant to understanding electron-phonon
driven superconductivity in SrTiO$_3$ \cite{SrTiO3First}, especially given its recently
uncovered one-dimensional nature \cite{SrTiO31D},} {\color{darkblue2}{in magic-angle graphene \cite{MAGraphene}, and layered MoS$_2$ \cite{MoS2}.}}

To validate our proposed new pathway to high-temperature superconductivity, a
detailed understanding of Peierls/SSH couplings and their interplay with Coulomb repulsion
at finite carrier concentrations is required, see \cite{Mixed_halffilling, ErezBerg_nonFermiSSH} for 
recent studies. 





All these
considerations indicate that the issue of phonon-mediated
high-temperature superconductivity must be revisited.

\begin{acknowledgments}
We thank Daniel Dessau and Mirko M\"oller for useful discussions. This work was supported by the Natural Sciences and Engineering Research Council of Canada (NSERC) (J.S., R.V.K. and M.B.) and the Stewart Blusson Quantum Matter Institute (SBQMI) (J.S. and M.B.).  J.S. is also supported by a visiting student fellowship at the Institute for Theoretical Atomic, Molecular, and Optical Physics (ITAMP) at Harvard University and the Smithsonian Astrophysical Observatory. M.C. appreciates access to the computing facilities of the DST-FIST (phase-II) project installed in the Department of Physics, Indian Institute of Technology (IIT), Kharagpur, India.
\end{acknowledgments}


\begin{thebibliography}{60}%
\makeatletter
\providecommand \@ifxundefined [1]{%
 \@ifx{#1\undefined}
}%
\providecommand \@ifnum [1]{%
 \ifnum #1\expandafter \@firstoftwo
 \else \expandafter \@secondoftwo
 \fi
}%
\providecommand \@ifx [1]{%
 \ifx #1\expandafter \@firstoftwo
 \else \expandafter \@secondoftwo
 \fi
}%
\providecommand \natexlab [1]{#1}%
\providecommand \enquote  [1]{``#1''}%
\providecommand \bibnamefont  [1]{#1}%
\providecommand \bibfnamefont [1]{#1}%
\providecommand \citenamefont [1]{#1}%
\providecommand \href@noop [0]{\@secondoftwo}%
\providecommand \href [0]{\begingroup \@sanitize@url \@href}%
\providecommand \@href[1]{\@@startlink{#1}\@@href}%
\providecommand \@@href[1]{\endgroup#1\@@endlink}%
\providecommand \@sanitize@url [0]{\catcode `\\12\catcode `\$12\catcode
  `\&12\catcode `\#12\catcode `\^12\catcode `\_12\catcode `\%12\relax}%
\providecommand \@@startlink[1]{}%
\providecommand \@@endlink[0]{}%
\providecommand \url  [0]{\begingroup\@sanitize@url \@url }%
\providecommand \@url [1]{\endgroup\@href {#1}{\urlprefix }}%
\providecommand \urlprefix  [0]{URL }%
\providecommand \Eprint [0]{\href }%
\providecommand \doibase [0]{http://dx.doi.org/}%
\providecommand \selectlanguage [0]{\@gobble}%
\providecommand \bibinfo  [0]{\@secondoftwo}%
\providecommand \bibfield  [0]{\@secondoftwo}%
\providecommand \translation [1]{[#1]}%
\providecommand \BibitemOpen [0]{}%
\providecommand \bibitemStop [0]{}%
\providecommand \bibitemNoStop [0]{.\EOS\space}%
\providecommand \EOS [0]{\spacefactor3000\relax}%
\providecommand \BibitemShut  [1]{\csname bibitem#1\endcsname}%
\let\auto@bib@innerbib\@empty
\bibitem [{\citenamefont {Kamerlingh-Onnes}(1911)}]{Firstsuperconductor}%
  \BibitemOpen
  \bibfield  {author} {\bibinfo {author} {\bibfnamefont {H.}~\bibnamefont
  {Kamerlingh-Onnes}},\ }\href@noop {} {\bibfield  {journal} {\bibinfo
  {journal} {Comm. Phys. Lab. Univ. Leiden}\,\ \bibinfo {pages} {124}}
  (\bibinfo {year} {1911})}\BibitemShut {NoStop}%
\bibitem [{\citenamefont {Bednorz}\ and\ \citenamefont
  {M{\"u}ller}(1986)}]{cuprate_super}%
  \BibitemOpen
  \bibfield  {author} {\bibinfo {author} {\bibfnamefont {J.~G.}\ \bibnamefont
  {Bednorz}}\ and\ \bibinfo {author} {\bibfnamefont {K.~A.}\ \bibnamefont
  {M{\"u}ller}},\ }\href@noop {} {\bibfield  {journal} {\bibinfo  {journal}
  {Z. Phys. B}\ }\textbf {\bibinfo {volume}
  {64}},\ \bibinfo {pages} {189} (\bibinfo {year} {1986})}\BibitemShut
  {NoStop}%
\bibitem [{\citenamefont {Kamihara}\ \emph {et~al.}(2008)\citenamefont
  {Kamihara}, \citenamefont {Watanabe}, \citenamefont {Hirano},\ and\
  \citenamefont {Hosono}}]{Ironbased_super1}%
  \BibitemOpen
  \bibfield  {author} {\bibinfo {author} {\bibfnamefont {Y.}~\bibnamefont
  {Kamihara}}, \bibinfo {author} {\bibfnamefont {T.}~\bibnamefont {Watanabe}},
  \bibinfo {author} {\bibfnamefont {M.}~\bibnamefont {Hirano}}, \ and\ \bibinfo
  {author} {\bibfnamefont {H.}~\bibnamefont {Hosono}},\ }\href {\doibase
  10.1021/ja800073m} {\bibfield  {journal} {\bibinfo  {journal} {J.
  Am. Chem. Soc.}\ }\textbf {\bibinfo {volume} {130}},\ \bibinfo
  {pages} {3296} (\bibinfo {year} {2008})}\BibitemShut {NoStop}%
\bibitem [{\citenamefont {Takahashi}\ \emph {et~al.}(2008)\citenamefont
  {Takahashi}, \citenamefont {Igawa}, \citenamefont {Arii}, \citenamefont
  {Kamihara}, \citenamefont {Hirano},\ and\ \citenamefont
  {Hosono}}]{Ironbased_super2}%
  \BibitemOpen
  \bibfield  {author} {\bibinfo {author} {\bibfnamefont {H.}~\bibnamefont
  {Takahashi}}, \bibinfo {author} {\bibfnamefont {K.}~\bibnamefont {Igawa}},
  \bibinfo {author} {\bibfnamefont {K.}~\bibnamefont {Arii}}, \bibinfo {author}
  {\bibfnamefont {Y.}~\bibnamefont {Kamihara}}, \bibinfo {author}
  {\bibfnamefont {M.}~\bibnamefont {Hirano}}, \ and\ \bibinfo {author}
  {\bibfnamefont {H.}~\bibnamefont {Hosono}},\ }\href
  {http://dx.doi.org/10.1038/nature06972} {\bibfield  {journal} {\bibinfo
  {journal} {Nature}\ }\textbf {\bibinfo {volume} {453}},\ \bibinfo {pages}
  {376 EP } (\bibinfo {year} {2008})}\BibitemShut {NoStop}%
\bibitem [{\citenamefont {Bardeen}\ \emph
  {et~al.}(1957{\natexlab{a}})\citenamefont {Bardeen}, \citenamefont {Cooper},\
  and\ \citenamefont {Schrieffer}}]{BCS1}%
  \BibitemOpen
  \bibfield  {author} {\bibinfo {author} {\bibfnamefont {J.}~\bibnamefont
  {Bardeen}}, \bibinfo {author} {\bibfnamefont {L.~N.}\ \bibnamefont {Cooper}},
  \ and\ \bibinfo {author} {\bibfnamefont {J.~R.}\ \bibnamefont {Schrieffer}},\
  }\href {\doibase 10.1103/PhysRev.106.162} {\bibfield  {journal} {\bibinfo
  {journal} {Phys. Rev.}\ }\textbf {\bibinfo {volume} {106}},\ \bibinfo {pages}
  {162} (\bibinfo {year} {1957}{\natexlab{a}})}\BibitemShut {NoStop}%
\bibitem [{\citenamefont {Bardeen}\ \emph
  {et~al.}(1957{\natexlab{b}})\citenamefont {Bardeen}, \citenamefont {Cooper},\
  and\ \citenamefont {Schrieffer}}]{BCS2}%
  \BibitemOpen
  \bibfield  {author} {\bibinfo {author} {\bibfnamefont {J.}~\bibnamefont
  {Bardeen}}, \bibinfo {author} {\bibfnamefont {L.~N.}\ \bibnamefont {Cooper}},
  \ and\ \bibinfo {author} {\bibfnamefont {J.~R.}\ \bibnamefont {Schrieffer}},\
  }\href {\doibase 10.1103/PhysRev.108.1175} {\bibfield  {journal} {\bibinfo
  {journal} {Phys. Rev.}\ }\textbf {\bibinfo {volume} {108}},\ \bibinfo {pages}
  {1175} (\bibinfo {year} {1957}{\natexlab{b}})}\BibitemShut {NoStop}%
\bibitem [{\citenamefont {Cooper}(1956)}]{CooperPair}%
  \BibitemOpen
  \bibfield  {author} {\bibinfo {author} {\bibfnamefont {L.~N.}\ \bibnamefont
  {Cooper}},\ }\href {\doibase 10.1103/PhysRev.104.1189} {\bibfield  {journal}
  {\bibinfo  {journal} {Phys. Rev.}\ }\textbf {\bibinfo {volume} {104}},\
  \bibinfo {pages} {1189} (\bibinfo {year} {1956})}\BibitemShut {NoStop}%
\bibitem [{\citenamefont {Landau}(1933)}]{Landau}%
  \BibitemOpen
  \bibfield  {author} {\bibinfo {author} {\bibfnamefont {L.}~\bibnamefont
  {Landau}},\ }\href@noop {} {\bibfield  {journal} {\bibinfo  {journal} {Phys.
  Z. Sowjetunion}\ }\textbf {\bibinfo {volume} {3}},\ \bibinfo {pages} {644}
  (\bibinfo {year} {1933})}\BibitemShut {NoStop}%
\bibitem [{\citenamefont {Landau}\ and\ \citenamefont
  {Pekar}(1948)}]{Landau_Pekar}%
  \BibitemOpen
  \bibfield  {author} {\bibinfo {author} {\bibfnamefont {L.}~\bibnamefont
  {Landau}}\ and\ \bibinfo {author} {\bibfnamefont {S.}~\bibnamefont {Pekar}},\
  }\href@noop {} {\bibfield  {journal} {\bibinfo  {journal} {J. Exp. Theor.
  Phys}\ }\textbf {\bibinfo {volume} {18}},\ \bibinfo {pages} {419} (\bibinfo
  {year} {1948})}\BibitemShut {NoStop}%
\bibitem [{\citenamefont {Fr{\"o}hlich}\ \emph {et~al.}(1950)\citenamefont
  {Fr{\"o}hlich}, \citenamefont {Pelzer},\ and\ \citenamefont
  {Zienau}}]{Froh1}%
  \BibitemOpen
  \bibfield  {author} {\bibinfo {author} {\bibfnamefont {H.}~\bibnamefont
  {Fr{\"o}hlich}}, \bibinfo {author} {\bibfnamefont {H.}~\bibnamefont
  {Pelzer}}, \ and\ \bibinfo {author} {\bibfnamefont {S.}~\bibnamefont
  {Zienau}},\ }\href@noop {} {\bibfield  {journal} {\bibinfo  {journal}
  {Philos. Mag.}\ }\textbf {\bibinfo {volume} {41}},\ \bibinfo {pages} {221}
  (\bibinfo {year} {1950})}\BibitemShut {NoStop}%
\bibitem [{\citenamefont {Tyablikov}(1952)}]{Holstein1}%
  \BibitemOpen
  \bibfield  {author} {\bibinfo {author} {\bibfnamefont {S.}~\bibnamefont
  {Tyablikov}},\ }\href@noop {} {\bibfield  {journal} {\bibinfo  {journal} {Zh.
  Eksp. Teor. Fiz}\ }\textbf {\bibinfo {volume} {23}},\ \bibinfo {pages} {381}
  (\bibinfo {year} {1952})}\BibitemShut {NoStop}%
\bibitem [{\citenamefont {Fr{\"o}hlich}(1954)}]{Froh2}%
  \BibitemOpen
  \bibfield  {author} {\bibinfo {author} {\bibfnamefont {H.}~\bibnamefont
  {Fr{\"o}hlich}},\ }\href@noop {} {\bibfield  {journal} {\bibinfo  {journal}
  {Adv. Phys.}\ }\textbf {\bibinfo {volume} {3}},\ \bibinfo {pages} {325}
  (\bibinfo {year} {1954})}\BibitemShut {NoStop}%
\bibitem [{\citenamefont {Holstein}(2000)}]{Holstein2}%
  \BibitemOpen
  \bibfield  {author} {\bibinfo {author} {\bibfnamefont {T.}~\bibnamefont
  {Holstein}},\ }\href {\doibase https://doi.org/10.1006/aphy.2000.6021}
  {\bibfield  {journal} {\bibinfo  {journal} {Ann. Phys.}\ }\textbf {\bibinfo
  {volume} {281}},\ \bibinfo {pages} {725 } (\bibinfo {year}
  {2000})}\BibitemShut {NoStop}%
\bibitem [{\citenamefont {Feynman}(1955)}]{Feynman1}%
  \BibitemOpen
  \bibfield  {author} {\bibinfo {author} {\bibfnamefont {R.~P.}\ \bibnamefont
  {Feynman}},\ }\href@noop {} {\bibfield  {journal} {\bibinfo  {journal} {Phys.
  Rev.}\ }\textbf {\bibinfo {volume} {97}},\ \bibinfo {pages} {660} (\bibinfo
  {year} {1955})}\BibitemShut {NoStop}%
\bibitem [{\citenamefont {Feynman}\ \emph {et~al.}(1962)\citenamefont
  {Feynman}, \citenamefont {Hellwarth}, \citenamefont {Iddings},\ and\
  \citenamefont {Platzman}}]{Feynman2}%
  \BibitemOpen
  \bibfield  {author} {\bibinfo {author} {\bibfnamefont {R.}~\bibnamefont
  {Feynman}}, \bibinfo {author} {\bibfnamefont {R.}~\bibnamefont {Hellwarth}},
  \bibinfo {author} {\bibfnamefont {C.}~\bibnamefont {Iddings}}, \ and\
  \bibinfo {author} {\bibfnamefont {P.}~\bibnamefont {Platzman}},\ }\href@noop
  {} {\bibfield  {journal} {\bibinfo  {journal} {Phys. Rev.}\ }\textbf
  {\bibinfo {volume} {127}},\ \bibinfo {pages} {1004} (\bibinfo {year}
  {1962})}\BibitemShut {NoStop}%
\bibitem [{\citenamefont {Prokof'ev}\ and\ \citenamefont
  {Svistunov}(1998{\natexlab{a}})}]{Prokofiev_FrohPolaron1}%
  \BibitemOpen
  \bibfield  {author} {\bibinfo {author} {\bibfnamefont {N.~V.}\ \bibnamefont
  {Prokof'ev}}\ and\ \bibinfo {author} {\bibfnamefont {B.~V.}\ \bibnamefont
  {Svistunov}},\ }\href {\doibase 10.1103/PhysRevLett.81.2514} {\bibfield
  {journal} {\bibinfo  {journal} {Phys. Rev. Lett.}\ }\textbf {\bibinfo
  {volume} {81}},\ \bibinfo {pages} {2514} (\bibinfo {year}
  {1998}{\natexlab{a}})}\BibitemShut {NoStop}%
\bibitem [{\citenamefont {Prokof'ev}\ and\ \citenamefont
  {Svistunov}(1998{\natexlab{b}})}]{Prokofiev_FrohPolaron2}%
  \BibitemOpen
  \bibfield  {author} {\bibinfo {author} {\bibfnamefont {N.~V.}\ \bibnamefont
  {Prokof'ev}}\ and\ \bibinfo {author} {\bibfnamefont {B.~V.}\ \bibnamefont
  {Svistunov}},\ }\href {\doibase 10.1103/PhysRevLett.81.2514} {\bibfield
  {journal} {\bibinfo  {journal} {Phys. Rev. Lett.}\ }\textbf {\bibinfo
  {volume} {81}},\ \bibinfo {pages} {2514} (\bibinfo {year}
  {1998}{\natexlab{b}})}\BibitemShut {NoStop}%
\bibitem [{\citenamefont {Kornilovitch}(1998)}]{Kornilovitch_polaron_general1}%
  \BibitemOpen
  \bibfield  {author} {\bibinfo {author} {\bibfnamefont {P.~E.}\ \bibnamefont
  {Kornilovitch}},\ }\href {\doibase 10.1103/PhysRevLett.81.5382} {\bibfield
  {journal} {\bibinfo  {journal} {Phys. Rev. Lett.}\ }\textbf {\bibinfo
  {volume} {81}},\ \bibinfo {pages} {5382} (\bibinfo {year}
  {1998})}\BibitemShut {NoStop}%
\bibitem [{\citenamefont {Bon\ifmmode~\check{c}\else \v{c}\fi{}a}\ \emph
  {et~al.}(1999)\citenamefont {Bon\ifmmode~\check{c}\else \v{c}\fi{}a},
  \citenamefont {Trugman},\ and\ \citenamefont {Batisti\ifmmode~\acute{c}\else
  \'{c}\fi{}}}]{Bonca_polaron1}%
  \BibitemOpen
  \bibfield  {author} {\bibinfo {author} {\bibfnamefont {J.}~\bibnamefont
  {Bon\ifmmode~\check{c}\else \v{c}\fi{}a}}, \bibinfo {author} {\bibfnamefont
  {S.~A.}\ \bibnamefont {Trugman}}, \ and\ \bibinfo {author} {\bibfnamefont
  {I.}~\bibnamefont {Batisti\ifmmode~\acute{c}\else \'{c}\fi{}}},\ }\href
  {\doibase 10.1103/PhysRevB.60.1633} {\bibfield  {journal} {\bibinfo
  {journal} {Phys. Rev. B}\ }\textbf {\bibinfo {volume} {60}},\ \bibinfo
  {pages} {1633} (\bibinfo {year} {1999})}\BibitemShut {NoStop}%
\bibitem [{\citenamefont {Ku}\ \emph {et~al.}(2002)\citenamefont {Ku},
  \citenamefont {Trugman},\ and\ \citenamefont {Bon\ifmmode~\check{c}\else
  \v{c}\fi{}a}}]{Bonca_polaron2}%
  \BibitemOpen
  \bibfield  {author} {\bibinfo {author} {\bibfnamefont {L.-C.}\ \bibnamefont
  {Ku}}, \bibinfo {author} {\bibfnamefont {S.~A.}\ \bibnamefont {Trugman}}, \
  and\ \bibinfo {author} {\bibfnamefont {J.}~\bibnamefont
  {Bon\ifmmode~\check{c}\else \v{c}\fi{}a}},\ }\href {\doibase
  10.1103/PhysRevB.65.174306} {\bibfield  {journal} {\bibinfo  {journal} {Phys.
  Rev. B}\ }\textbf {\bibinfo {volume} {65}},\ \bibinfo {pages} {174306}
  (\bibinfo {year} {2002})}\BibitemShut {NoStop}%
\bibitem [{\citenamefont {Hague}\ \emph {et~al.}(2006)\citenamefont {Hague},
  \citenamefont {Kornilovitch}, \citenamefont {Alexandrov},\ and\ \citenamefont
  {Samson}}]{Kornilovitch_polaron_dimensionality2}%
  \BibitemOpen
  \bibfield  {author} {\bibinfo {author} {\bibfnamefont {J.~P.}\ \bibnamefont
  {Hague}}, \bibinfo {author} {\bibfnamefont {P.~E.}\ \bibnamefont
  {Kornilovitch}}, \bibinfo {author} {\bibfnamefont {A.~S.}\ \bibnamefont
  {Alexandrov}}, \ and\ \bibinfo {author} {\bibfnamefont {J.~H.}\ \bibnamefont
  {Samson}},\ }\href {\doibase 10.1103/PhysRevB.73.054303} {\bibfield
  {journal} {\bibinfo  {journal} {Phys. Rev. B}\ }\textbf {\bibinfo {volume}
  {73}},\ \bibinfo {pages} {054303} (\bibinfo {year} {2006})}\BibitemShut
  {NoStop}%
\bibitem [{\citenamefont {Goodvin}\ \emph {et~al.}(2006)\citenamefont
  {Goodvin}, \citenamefont {Berciu},\ and\ \citenamefont
  {Sawatzky}}]{Holstein_MA}%
  \BibitemOpen
  \bibfield  {author} {\bibinfo {author} {\bibfnamefont {G.~L.}\ \bibnamefont
  {Goodvin}}, \bibinfo {author} {\bibfnamefont {M.}~\bibnamefont {Berciu}}, \
  and\ \bibinfo {author} {\bibfnamefont {G.~A.}\ \bibnamefont {Sawatzky}},\
  }\href {\doibase 10.1103/PhysRevB.74.245104} {\bibfield  {journal} {\bibinfo
  {journal} {Phys. Rev. B}\ }\textbf {\bibinfo {volume} {74}},\ \bibinfo
  {pages} {245104} (\bibinfo {year} {2006})}\BibitemShut {NoStop}%
\bibitem [{\citenamefont {Chakraverty}\ \emph {et~al.}(1998)\citenamefont
  {Chakraverty}, \citenamefont {Ranninger},\ and\ \citenamefont
  {Feinberg}}]{ImpossibleBipolaronicSuperconductivity}%
  \BibitemOpen
  \bibfield  {author} {\bibinfo {author} {\bibfnamefont {B.~K.}\ \bibnamefont
  {Chakraverty}}, \bibinfo {author} {\bibfnamefont {J.}~\bibnamefont
  {Ranninger}}, \ and\ \bibinfo {author} {\bibfnamefont {D.}~\bibnamefont
  {Feinberg}},\ }\href {\doibase 10.1103/PhysRevLett.81.433} {\bibfield
  {journal} {\bibinfo  {journal} {Phys. Rev. Lett.}\ }\textbf {\bibinfo
  {volume} {81}},\ \bibinfo {pages} {433} (\bibinfo {year} {1998})}\BibitemShut
  {NoStop}%
\bibitem [{Note1()}]{Note1}%
  \BibitemOpen
  \bibinfo {note} { {\color{darkblue2}{High-$T_c$ bipolaronic superconductivity may be possible in
  very special circumstances like in Ref. \cite
  {HagueSuperLight}}}}\BibitemShut {NoStop}%
\bibitem [{\citenamefont {Bari\ifmmode \check{s}\else
  \v{s}\fi{}i\ifmmode~\acute{c}\else \'{c}\fi{}}\ \emph
  {et~al.}(1970)\citenamefont {Bari\ifmmode \check{s}\else
  \v{s}\fi{}i\ifmmode~\acute{c}\else \'{c}\fi{}}, \citenamefont {Labb\'e},\
  and\ \citenamefont {Friedel}}]{SSH1a}%
  \BibitemOpen
  \bibfield  {author} {\bibinfo {author} {\bibfnamefont {S.}~\bibnamefont
  {Bari\ifmmode \check{s}\else \v{s}\fi{}i\ifmmode~\acute{c}\else \'{c}\fi{}}},
  \bibinfo {author} {\bibfnamefont {J.}~\bibnamefont {Labb\'e}}, \ and\
  \bibinfo {author} {\bibfnamefont {J.}~\bibnamefont {Friedel}},\ }\href
  {\doibase 10.1103/PhysRevLett.25.919} {\bibfield  {journal} {\bibinfo
  {journal} {Phys. Rev. Lett.}\ }\textbf {\bibinfo {volume} {25}},\ \bibinfo
  {pages} {919} (\bibinfo {year} {1970})}\BibitemShut {NoStop}%
\bibitem [{\citenamefont {Bari\ifmmode \check{s}\else
  \v{s}\fi{}i\ifmmode~\acute{c}\else \'{c}\fi{}}(1972{\natexlab{a}})}]{SSH1b}%
  \BibitemOpen
  \bibfield  {author} {\bibinfo {author} {\bibfnamefont {S.}~\bibnamefont
  {Bari\ifmmode \check{s}\else \v{s}\fi{}i\ifmmode~\acute{c}\else
  \'{c}\fi{}}},\ }\href {\doibase 10.1103/PhysRevB.5.932} {\bibfield  {journal}
  {\bibinfo  {journal} {Phys. Rev. B}\ }\textbf {\bibinfo {volume} {5}},\
  \bibinfo {pages} {932} (\bibinfo {year} {1972}{\natexlab{a}})}\BibitemShut
  {NoStop}%
\bibitem [{\citenamefont {Bari\ifmmode \check{s}\else
  \v{s}\fi{}i\ifmmode~\acute{c}\else \'{c}\fi{}}(1972{\natexlab{b}})}]{SSH1c}%
  \BibitemOpen
  \bibfield  {author} {\bibinfo {author} {\bibfnamefont {S.}~\bibnamefont
  {Bari\ifmmode \check{s}\else \v{s}\fi{}i\ifmmode~\acute{c}\else
  \'{c}\fi{}}},\ }\href {\doibase 10.1103/PhysRevB.5.941} {\bibfield  {journal}
  {\bibinfo  {journal} {Phys. Rev. B}\ }\textbf {\bibinfo {volume} {5}},\
  \bibinfo {pages} {941} (\bibinfo {year} {1972}{\natexlab{b}})}\BibitemShut
  {NoStop}%
\bibitem [{\citenamefont {Su}\ \emph {et~al.}(1979)\citenamefont {Su},
  \citenamefont {Schrieffer},\ and\ \citenamefont {Heeger}}]{SSH2a}%
  \BibitemOpen
  \bibfield  {author} {\bibinfo {author} {\bibfnamefont {W.~P.}\ \bibnamefont
  {Su}}, \bibinfo {author} {\bibfnamefont {J.~R.}\ \bibnamefont {Schrieffer}},
  \ and\ \bibinfo {author} {\bibfnamefont {A.~J.}\ \bibnamefont {Heeger}},\
  }\href {\doibase 10.1103/PhysRevLett.42.1698} {\bibfield  {journal} {\bibinfo
   {journal} {Phys. Rev. Lett.}\ }\textbf {\bibinfo {volume} {42}},\ \bibinfo
  {pages} {1698} (\bibinfo {year} {1979})}\BibitemShut {NoStop}%
\bibitem [{\citenamefont {Heeger}\ \emph {et~al.}(1988)\citenamefont {Heeger},
  \citenamefont {Kivelson}, \citenamefont {Schrieffer},\ and\ \citenamefont
  {Su}}]{SSH2b}%
  \BibitemOpen
  \bibfield  {author} {\bibinfo {author} {\bibfnamefont {A.~J.}\ \bibnamefont
  {Heeger}}, \bibinfo {author} {\bibfnamefont {S.}~\bibnamefont {Kivelson}},
  \bibinfo {author} {\bibfnamefont {J.~R.}\ \bibnamefont {Schrieffer}}, \ and\
  \bibinfo {author} {\bibfnamefont {W.~P.}\ \bibnamefont {Su}},\ }\href
  {\doibase 10.1103/RevModPhys.60.781} {\bibfield  {journal} {\bibinfo
  {journal} {Rev. Mod. Phys.}\ }\textbf {\bibinfo {volume} {60}},\ \bibinfo
  {pages} {781} (\bibinfo {year} {1988})}\BibitemShut {NoStop}%
\bibitem [{\citenamefont {Marchand}\ \emph {et~al.}(2010)\citenamefont
  {Marchand}, \citenamefont {De~Filippis}, \citenamefont {Cataudella},
  \citenamefont {Berciu}, \citenamefont {Nagaosa}, \citenamefont {Prokof'ev},
  \citenamefont {Mishchenko},\ and\ \citenamefont {Stamp}}]{Dominic}%
  \BibitemOpen
  \bibfield  {author} {\bibinfo {author} {\bibfnamefont {D.~J.~J.}\
  \bibnamefont {Marchand}}, \bibinfo {author} {\bibfnamefont {G.}~\bibnamefont
  {De~Filippis}}, \bibinfo {author} {\bibfnamefont {V.}~\bibnamefont
  {Cataudella}}, \bibinfo {author} {\bibfnamefont {M.}~\bibnamefont {Berciu}},
  \bibinfo {author} {\bibfnamefont {N.}~\bibnamefont {Nagaosa}}, \bibinfo
  {author} {\bibfnamefont {N.~V.}\ \bibnamefont {Prokof'ev}}, \bibinfo {author}
  {\bibfnamefont {A.~S.}\ \bibnamefont {Mishchenko}}, \ and\ \bibinfo {author}
  {\bibfnamefont {P.~C.~E.}\ \bibnamefont {Stamp}},\ }\href {\doibase
  10.1103/PhysRevLett.105.266605} {\bibfield  {journal} {\bibinfo  {journal}
  {Phys. Rev. Lett.}\ }\textbf {\bibinfo {volume} {105}},\ \bibinfo {pages}
  {266605} (\bibinfo {year} {2010})}\BibitemShut {NoStop}%
\bibitem [{\citenamefont {Bon\ifmmode~\check{c}\else \v{c}\fi{}a}\ \emph
  {et~al.}(2000)\citenamefont {Bon\ifmmode~\check{c}\else \v{c}\fi{}a},
  \citenamefont {Katra\ifmmode~\check{s}\else \v{s}\fi{}nik},\ and\
  \citenamefont {Trugman}}]{Bonca_bipolaron1}%
  \BibitemOpen
  \bibfield  {author} {\bibinfo {author} {\bibfnamefont {J.}~\bibnamefont
  {Bon\ifmmode~\check{c}\else \v{c}\fi{}a}}, \bibinfo {author} {\bibfnamefont
  {T.}~\bibnamefont {Katra\ifmmode~\check{s}\else \v{s}\fi{}nik}}, \ and\
  \bibinfo {author} {\bibfnamefont {S.~A.}\ \bibnamefont {Trugman}},\ }\href
  {\doibase 10.1103/PhysRevLett.84.3153} {\bibfield  {journal} {\bibinfo
  {journal} {Phys. Rev. Lett.}\ }\textbf {\bibinfo {volume} {84}},\ \bibinfo
  {pages} {3153} (\bibinfo {year} {2000})}\BibitemShut {NoStop}%
\bibitem [{\citenamefont {Bon\ifmmode~\check{c}\else \v{c}\fi{}a}\ and\
  \citenamefont {Trugman}(2001)}]{Bonca_bipolaron2}%
  \BibitemOpen
  \bibfield  {author} {\bibinfo {author} {\bibfnamefont {J.}~\bibnamefont
  {Bon\ifmmode~\check{c}\else \v{c}\fi{}a}}\ and\ \bibinfo {author}
  {\bibfnamefont {S.~A.}\ \bibnamefont {Trugman}},\ }\href {\doibase
  10.1103/PhysRevB.64.094507} {\bibfield  {journal} {\bibinfo  {journal} {Phys.
  Rev. B}\ }\textbf {\bibinfo {volume} {64}},\ \bibinfo {pages} {094507}
  (\bibinfo {year} {2001})}\BibitemShut {NoStop}%
\bibitem [{\citenamefont {Chakraborty}\ \emph {et~al.}(2012)\citenamefont
  {Chakraborty}, \citenamefont {Min}, \citenamefont {Chakrabarti},\ and\
  \citenamefont {Das}}]{Monodeep_bipolaron}%
  \BibitemOpen
  \bibfield  {author} {\bibinfo {author} {\bibfnamefont {M.}~\bibnamefont
  {Chakraborty}}, \bibinfo {author} {\bibfnamefont {B.~I.}\ \bibnamefont
  {Min}}, \bibinfo {author} {\bibfnamefont {A.}~\bibnamefont {Chakrabarti}}, \
  and\ \bibinfo {author} {\bibfnamefont {A.~N.}\ \bibnamefont {Das}},\ }\href
  {\doibase 10.1103/PhysRevB.85.245127} {\bibfield  {journal} {\bibinfo
  {journal} {Phys. Rev. B}\ }\textbf {\bibinfo {volume} {85}},\ \bibinfo
  {pages} {245127} (\bibinfo {year} {2012})}\BibitemShut {NoStop}%
\bibitem [{\citenamefont {Berciu}(2006)}]{Mona_MA}%
  \BibitemOpen
  \bibfield  {author} {\bibinfo {author} {\bibfnamefont {M.}~\bibnamefont
  {Berciu}},\ }\href {\doibase 10.1103/PhysRevLett.97.036402} {\bibfield
  {journal} {\bibinfo  {journal} {Phys. Rev. Lett.}\ }\textbf {\bibinfo
  {volume} {97}},\ \bibinfo {pages} {036402} (\bibinfo {year}
  {2006})}\BibitemShut {NoStop}%
\bibitem [{\citenamefont {Sous}\ \emph {et~al.}(2017)\citenamefont {Sous},
  \citenamefont {Chakraborty}, \citenamefont {Adolphs}, \citenamefont {Krems},\
  and\ \citenamefont {Berciu}}]{Repulsive}%
  \BibitemOpen
  \bibfield  {author} {\bibinfo {author} {\bibfnamefont {J.}~\bibnamefont
  {Sous}}, \bibinfo {author} {\bibfnamefont {M.}~\bibnamefont {Chakraborty}},
  \bibinfo {author} {\bibfnamefont {C.~P.~J.}\ \bibnamefont {Adolphs}},
  \bibinfo {author} {\bibfnamefont {R.~V.}\ \bibnamefont {Krems}}, \ and\
  \bibinfo {author} {\bibfnamefont {M.}~\bibnamefont {Berciu}},\ }\href@noop {}
  {\bibfield  {journal} {\bibinfo  {journal} {Sci. Rep.}\ }\textbf {\bibinfo
  {volume} {7}},\ \bibinfo {pages} {1169} (\bibinfo {year} {2017})}\BibitemShut
  {NoStop}%
\bibitem [{\citenamefont {Sous}\ \emph {et~al.}()\citenamefont {Sous} \emph
  {et~al.}}]{SousLongPaper}%
  \BibitemOpen
  \bibfield  {author} {\bibinfo {author} {\bibfnamefont {J.}~\bibnamefont
  {Sous}} \emph {et~al.},\ }\href@noop {} {\bibinfo  {journal} {in
  preparation}\ }\BibitemShut {NoStop}%
\bibitem [{\citenamefont {Takahashi}(1977)}]{Projection_technique}%
  \BibitemOpen
\bibfield  {journal} {  }\bibfield  {author} {\bibinfo {author} {\bibfnamefont
  {M.}~\bibnamefont {Takahashi}},\ }\href@noop {} {\bibfield  {journal}
  {\bibinfo  {journal} {J. Phys. C: Solid State Phys.}\ }\textbf {\bibinfo
  {volume} {10}},\ \bibinfo {pages} {1289} (\bibinfo {year}
  {1977})}\BibitemShut {NoStop}%
\bibitem [{\citenamefont {Berciu}(2011)}]{MBmp}%
  \BibitemOpen
  \bibfield  {author} {\bibinfo {author} {\bibfnamefont {M.}~\bibnamefont
  {Berciu}},\ }\href {\doibase 10.1103/PhysRevLett.107.246403} {\bibfield
  {journal} {\bibinfo  {journal} {Phys. Rev. Lett.}\ }\textbf {\bibinfo
  {volume} {107}},\ \bibinfo {pages} {246403} (\bibinfo {year}
  {2011})}\BibitemShut {NoStop}%
{\color{darkblue2}{
  \bibitem  [{\citenamefont {Hague}\ \emph {et~al.}(2007)\citenamefont {Hague},
  \citenamefont {Kornilovitch}, \citenamefont {Samson},\ and\ \citenamefont
  {Alexandrov}}]{HagueSuperLight}%
  \BibitemOpen
  \bibfield  {author} {\bibinfo {author} {\bibfnamefont {J.~P.}\ \bibnamefont
  {Hague}}, \bibinfo {author} {\bibfnamefont {P.~E.}\ \bibnamefont
  {Kornilovitch}}, \bibinfo {author} {\bibfnamefont {J.~H.}\ \bibnamefont
  {Samson}}, \ and\ \bibinfo {author} {\bibfnamefont {A.~S.}\ \bibnamefont
  {Alexandrov}},\ }\href {\doibase 10.1103/PhysRevLett.98.037002} {\bibfield
  {journal} {\bibinfo  {journal} {Phys. Rev. Lett.}\ }\textbf {\bibinfo
  {volume} {98}},\ \bibinfo {pages} {037002} (\bibinfo {year}
  {2007})} \BibitemShut {NoStop}%
  }}
\bibitem [{\citenamefont {De~Sio}\ \emph {et~al.}(2016)\citenamefont {De~Sio},
  \citenamefont {Troiani}, \citenamefont {Maiuri}, \citenamefont {R{\'e}hault},
  \citenamefont {Sommer}, \citenamefont {Lim}, \citenamefont {Huelga},
  \citenamefont {Plenio}, \citenamefont {Rozzi}, \citenamefont {Cerullo},
  \citenamefont {Molinari},\ and\ \citenamefont {Lienau}}]{SSH_Polymers}%
  \BibitemOpen
  \bibfield  {author} {\bibinfo {author} {\bibfnamefont {A.}~\bibnamefont
  {De~Sio}}, \bibinfo {author} {\bibfnamefont {F.}~\bibnamefont {Troiani}},
  \bibinfo {author} {\bibfnamefont {M.}~\bibnamefont {Maiuri}}, \bibinfo
  {author} {\bibfnamefont {J.}~\bibnamefont {R{\'e}hault}}, \bibinfo {author}
  {\bibfnamefont {E.}~\bibnamefont {Sommer}}, \bibinfo {author} {\bibfnamefont
  {J.}~\bibnamefont {Lim}}, \bibinfo {author} {\bibfnamefont {S.~F.}\
  \bibnamefont {Huelga}}, \bibinfo {author} {\bibfnamefont {M.~B.}\
  \bibnamefont {Plenio}}, \bibinfo {author} {\bibfnamefont {C.~A.}\
  \bibnamefont {Rozzi}}, \bibinfo {author} {\bibfnamefont {G.}~\bibnamefont
  {Cerullo}}, \bibinfo {author} {\bibfnamefont {E.}~\bibnamefont {Molinari}}, \
  and\ \bibinfo {author} {\bibfnamefont {C.}~\bibnamefont {Lienau}},\ }\href
  {http://dx.doi.org/10.1038/ncomms13742} {\bibfield  {journal} {\bibinfo
  {journal} {Nat. Commun.}\ }\textbf {\bibinfo {volume} {7}},\ \bibinfo {pages}
  {13742} (\bibinfo {year} {2016})}\BibitemShut {NoStop}%
\bibitem [{\citenamefont {Nelson}\ \emph {et~al.}(1998)\citenamefont {Nelson},
  \citenamefont {Lin}, \citenamefont {Gundlach},\ and\ \citenamefont
  {Jackson}}]{Organic_SSH}%
  \BibitemOpen
  \bibfield  {author} {\bibinfo {author} {\bibfnamefont {S.}~\bibnamefont
  {Nelson}}, \bibinfo {author} {\bibfnamefont {Y.-Y.}\ \bibnamefont {Lin}},
  \bibinfo {author} {\bibfnamefont {D.}~\bibnamefont {Gundlach}}, \ and\
  \bibinfo {author} {\bibfnamefont {T.}~\bibnamefont {Jackson}},\ }\href@noop
  {} {\bibfield  {journal} {\bibinfo  {journal} {Appl. Phys. Lett.}\ }\textbf
  {\bibinfo {volume} {72}},\ \bibinfo {pages} {1854} (\bibinfo {year}
  {1998})}\BibitemShut {NoStop}%
\bibitem [{\citenamefont {Troisi}\ and\ \citenamefont
  {Orlandi}(2006)}]{Triosi1}%
  \BibitemOpen
  \bibfield  {author} {\bibinfo {author} {\bibfnamefont {A.}~\bibnamefont
  {Troisi}}\ and\ \bibinfo {author} {\bibfnamefont {G.}~\bibnamefont
  {Orlandi}},\ }\href {\doibase 10.1103/PhysRevLett.96.086601} {\bibfield
  {journal} {\bibinfo  {journal} {Phys. Rev. Lett.}\ }\textbf {\bibinfo
  {volume} {96}},\ \bibinfo {pages} {086601} (\bibinfo {year}
  {2006})}\BibitemShut {NoStop}%
\bibitem [{\citenamefont {De~Filippis}\ \emph {et~al.}(2015)\citenamefont
  {De~Filippis}, \citenamefont {Cataudella}, \citenamefont {Mishchenko},
  \citenamefont {Nagaosa}, \citenamefont {Fierro},\ and\ \citenamefont
  {de~Candia}}]{Candia}%
  \BibitemOpen
  \bibfield  {author} {\bibinfo {author} {\bibfnamefont {G.}~\bibnamefont
  {De~Filippis}}, \bibinfo {author} {\bibfnamefont {V.}~\bibnamefont
  {Cataudella}}, \bibinfo {author} {\bibfnamefont {A.~S.}\ \bibnamefont
  {Mishchenko}}, \bibinfo {author} {\bibfnamefont {N.}~\bibnamefont {Nagaosa}},
  \bibinfo {author} {\bibfnamefont {A.}~\bibnamefont {Fierro}}, \ and\ \bibinfo
  {author} {\bibfnamefont {A.}~\bibnamefont {de~Candia}},\ }\href {\doibase
  10.1103/PhysRevLett.114.086601} {\bibfield  {journal} {\bibinfo  {journal}
  {Phys. Rev. Lett.}\ }\textbf {\bibinfo {volume} {114}},\ \bibinfo {pages}
  {086601} (\bibinfo {year} {2015})}\BibitemShut {NoStop}%
\bibitem [{\citenamefont {Fratini}\ \emph {et~al.}(2017)\citenamefont
  {Fratini}, \citenamefont {Ciuchi}, \citenamefont {Mayou}, \citenamefont
  {de~Laissardiere},\ and\ \citenamefont {Troisi}}]{Triosi2}%
  \BibitemOpen
  \bibfield  {author} {\bibinfo {author} {\bibfnamefont {S.}~\bibnamefont
  {Fratini}}, \bibinfo {author} {\bibfnamefont {S.}~\bibnamefont {Ciuchi}},
  \bibinfo {author} {\bibfnamefont {D.}~\bibnamefont {Mayou}}, \bibinfo
  {author} {\bibfnamefont {G.~T.}\ \bibnamefont {de~Laissardiere}}, \ and\
  \bibinfo {author} {\bibfnamefont {A.}~\bibnamefont {Troisi}},\ }\href
  {http://dx.doi.org/10.1038/nmat4970} {\bibfield  {journal} {\bibinfo
  {journal} {Nat. Mater.}\ }\textbf {\bibinfo {volume} {16}},\ \bibinfo {pages}
  {998} (\bibinfo {year} {2017})}\BibitemShut {NoStop}%
\bibitem [{\citenamefont {Pearton}\ \emph {et~al.}(2004)\citenamefont
  {Pearton}, \citenamefont {Heo}, \citenamefont {Ivill}, \citenamefont
  {Norton},\ and\ \citenamefont {Steiner}}]{SemiconductorOxides_review}%
  \BibitemOpen
  \bibfield  {author} {\bibinfo {author} {\bibfnamefont {S.~J.}\ \bibnamefont
  {Pearton}}, \bibinfo {author} {\bibfnamefont {W.~H.}\ \bibnamefont {Heo}},
  \bibinfo {author} {\bibfnamefont {M.}~\bibnamefont {Ivill}}, \bibinfo
  {author} {\bibfnamefont {D.~P.}\ \bibnamefont {Norton}}, \ and\ \bibinfo
  {author} {\bibfnamefont {T.}~\bibnamefont {Steiner}},\ }\href
  {http://stacks.iop.org/0268-1242/19/i=10/a=R01} {\bibfield  {journal}
  {\bibinfo  {journal} {Semicond. Sci. Technol.}\ }\textbf {\bibinfo {volume}
  {19}},\ \bibinfo {pages} {R59} (\bibinfo {year} {2004})}\BibitemShut
  {NoStop}%
\bibitem [{\citenamefont {Johnston}\ \emph {et~al.}(2014)\citenamefont
  {Johnston}, \citenamefont {Mukherjee}, \citenamefont {Elfimov}, \citenamefont
  {Berciu},\ and\ \citenamefont {Sawatzky}}]{Steve_Berciu_Sawatzky}%
  \BibitemOpen
  \bibfield  {author} {\bibinfo {author} {\bibfnamefont {S.}~\bibnamefont
  {Johnston}}, \bibinfo {author} {\bibfnamefont {A.}~\bibnamefont {Mukherjee}},
  \bibinfo {author} {\bibfnamefont {I.}~\bibnamefont {Elfimov}}, \bibinfo
  {author} {\bibfnamefont {M.}~\bibnamefont {Berciu}}, \ and\ \bibinfo {author}
  {\bibfnamefont {G.~A.}\ \bibnamefont {Sawatzky}},\ }\href {\doibase
  10.1103/PhysRevLett.112.106404} {\bibfield  {journal} {\bibinfo  {journal}
  {Phys. Rev. Lett.}\ }\textbf {\bibinfo {volume} {112}},\ \bibinfo {pages}
  {106404} (\bibinfo {year} {2014})}\BibitemShut {NoStop}%
\bibitem [{\citenamefont {Herrera}\ and\ \citenamefont
  {Krems}(2011)}]{polar-molecules1}%
  \BibitemOpen
  \bibfield  {author} {\bibinfo {author} {\bibfnamefont {F.}~\bibnamefont
  {Herrera}}\ and\ \bibinfo {author} {\bibfnamefont {R.~V.}\ \bibnamefont
  {Krems}},\ }\href {\doibase 10.1103/PhysRevA.84.051401} {\bibfield  {journal}
  {\bibinfo  {journal} {Phys. Rev. A}\ }\textbf {\bibinfo {volume} {84}},\
  \bibinfo {pages} {051401} (\bibinfo {year} {2011})}\BibitemShut {NoStop}%
\bibitem [{\citenamefont {Herrera}\ \emph {et~al.}(2013)\citenamefont
  {Herrera}, \citenamefont {Madison}, \citenamefont {Krems},\ and\
  \citenamefont {Berciu}}]{polar-molecules2}%
  \BibitemOpen
  \bibfield  {author} {\bibinfo {author} {\bibfnamefont {F.}~\bibnamefont
  {Herrera}}, \bibinfo {author} {\bibfnamefont {K.~W.}\ \bibnamefont
  {Madison}}, \bibinfo {author} {\bibfnamefont {R.~V.}\ \bibnamefont {Krems}},
  \ and\ \bibinfo {author} {\bibfnamefont {M.}~\bibnamefont {Berciu}},\ }\href
  {\doibase 10.1103/PhysRevLett.110.223002} {\bibfield  {journal} {\bibinfo
  {journal} {Phys. Rev. Lett.}\ }\textbf {\bibinfo {volume} {110}},\ \bibinfo
  {pages} {223002} (\bibinfo {year} {2013})}\BibitemShut {NoStop}%
\bibitem [{\citenamefont {Hague}\ and\ \citenamefont
  {MacCormick}(2012)}]{rydbergs3}%
  \BibitemOpen
  \bibfield  {author} {\bibinfo {author} {\bibfnamefont {J.~P.}\ \bibnamefont
  {Hague}}\ and\ \bibinfo {author} {\bibfnamefont {C.}~\bibnamefont
  {MacCormick}},\ }\href {http://stacks.iop.org/1367-2630/14/i=3/a=033019}
  {\bibfield  {journal} {\bibinfo  {journal} {New J. Phys.}\ }\textbf {\bibinfo
  {volume} {14}},\ \bibinfo {pages} {033019} (\bibinfo {year}
  {2012})}\BibitemShut {NoStop}%
\bibitem [{\citenamefont {Mostame}\ \emph {et~al.}(2012)\citenamefont
  {Mostame}, \citenamefont {Rebentrost}, \citenamefont {Eisfeld}, \citenamefont
  {Kerman}, \citenamefont {Tsomokos},\ and\ \citenamefont
  {Aspuru-Guzik}}]{d-wave1}%
  \BibitemOpen
  \bibfield  {author} {\bibinfo {author} {\bibfnamefont {S.}~\bibnamefont
  {Mostame}}, \bibinfo {author} {\bibfnamefont {P.}~\bibnamefont {Rebentrost}},
  \bibinfo {author} {\bibfnamefont {A.}~\bibnamefont {Eisfeld}}, \bibinfo
  {author} {\bibfnamefont {A.~J.}\ \bibnamefont {Kerman}}, \bibinfo {author}
  {\bibfnamefont {D.~I.}\ \bibnamefont {Tsomokos}}, \ and\ \bibinfo {author}
  {\bibfnamefont {A.}~\bibnamefont {Aspuru-Guzik}},\ }\href
  {http://stacks.iop.org/1367-2630/14/i=10/a=105013} {\bibfield  {journal}
  {\bibinfo  {journal} {New J. Phys.}\ }\textbf {\bibinfo {volume} {14}},\
  \bibinfo {pages} {105013} (\bibinfo {year} {2012})}\BibitemShut {NoStop}%
\bibitem [{\citenamefont {Stojanovi\ifmmode~\acute{c}\else \'{c}\fi{}}\ \emph
  {et~al.}(2014)\citenamefont {Stojanovi\ifmmode~\acute{c}\else \'{c}\fi{}},
  \citenamefont {Vanevi\ifmmode~\acute{c}\else \'{c}\fi{}}, \citenamefont
  {Demler},\ and\ \citenamefont {Tian}}]{d-wave4}%
  \BibitemOpen
  \bibfield  {author} {\bibinfo {author} {\bibfnamefont {V.~M.}\ \bibnamefont
  {Stojanovi\ifmmode~\acute{c}\else \'{c}\fi{}}}, \bibinfo {author}
  {\bibfnamefont {M.}~\bibnamefont {Vanevi\ifmmode~\acute{c}\else \'{c}\fi{}}},
  \bibinfo {author} {\bibfnamefont {E.}~\bibnamefont {Demler}}, \ and\ \bibinfo
  {author} {\bibfnamefont {L.}~\bibnamefont {Tian}},\ }\href {\doibase
  10.1103/PhysRevB.89.144508} {\bibfield  {journal} {\bibinfo  {journal} {Phys.
  Rev. B}\ }\textbf {\bibinfo {volume} {89}},\ \bibinfo {pages} {144508}
  (\bibinfo {year} {2014})}\BibitemShut {NoStop}%
\bibitem [{\citenamefont {Lan}\ and\ \citenamefont
  {Lobo}(2014)}]{Lobo_SSHfermionicsimulator}%
  \BibitemOpen
  \bibfield  {author} {\bibinfo {author} {\bibfnamefont {Z.}~\bibnamefont
  {Lan}}\ and\ \bibinfo {author} {\bibfnamefont {C.}~\bibnamefont {Lobo}},\
  }\href {\doibase 10.1103/PhysRevA.90.033627} {\bibfield  {journal} {\bibinfo
  {journal} {Phys. Rev. A}\ }\textbf {\bibinfo {volume} {90}},\ \bibinfo
  {pages} {033627} (\bibinfo {year} {2014})}\BibitemShut {NoStop}%
\bibitem [{\citenamefont {Wang}\ \emph {et~al.}(2017)\citenamefont {Wang},
  \citenamefont {Gao}, \citenamefont {Huang},\ and\ \citenamefont
  {Chen}}]{RecordHighTcsuperconductivity1}%
  \BibitemOpen
  \bibfield  {author} {\bibinfo {author} {\bibfnamefont {R.-S.}\ \bibnamefont
  {Wang}}, \bibinfo {author} {\bibfnamefont {Y.}~\bibnamefont {Gao}}, \bibinfo
  {author} {\bibfnamefont {Z.-B.}\ \bibnamefont {Huang}}, \ and\ \bibinfo
  {author} {\bibfnamefont {X.-J.}\ \bibnamefont {Chen}},\ }\href@noop {}
  {\bibfield  {journal} {\bibinfo  {journal} {arXiv preprint arXiv:1703.06641}\
  } (\bibinfo {year} {2017})}\BibitemShut {NoStop}%
\bibitem [{\citenamefont {Li}\ \emph {et~al.}(2018)\citenamefont {Li},
  \citenamefont {Zhou}, \citenamefont {Parham}, \citenamefont {Nummy},
  \citenamefont {Griffith}, \citenamefont {Gordon}, \citenamefont
  {Chronister},\ and\ \citenamefont {Dessau}}]{RecordHighTcsuperconductivity2}%
  \BibitemOpen
  \bibfield  {author} {\bibinfo {author} {\bibfnamefont {H.}~\bibnamefont
  {Li}}, \bibinfo {author} {\bibfnamefont {X.}~\bibnamefont {Zhou}}, \bibinfo
  {author} {\bibfnamefont {S.}~\bibnamefont {Parham}}, \bibinfo {author}
  {\bibfnamefont {T.}~\bibnamefont {Nummy}}, \bibinfo {author} {\bibfnamefont
  {J.}~\bibnamefont {Griffith}}, \bibinfo {author} {\bibfnamefont
  {K.}~\bibnamefont {Gordon}}, \bibinfo {author} {\bibfnamefont {E.~L.}\
  \bibnamefont {Chronister}}, \ and\ \bibinfo {author} {\bibfnamefont
  {D.}~\bibnamefont {Dessau}},\ }\href@noop {} {\bibfield  {journal} {\bibinfo
  {journal} {arXiv preprint arXiv:1704.04230}\ } (\bibinfo {year}
  {2018})}\BibitemShut {NoStop}%
\bibitem [{\citenamefont {Schooley}\ \emph {et~al.}(1964)\citenamefont
  {Schooley}, \citenamefont {Hosler},\ and\ \citenamefont
  {Cohen}}]{SrTiO3First}%
  \BibitemOpen
  \bibfield  {author} {\bibinfo {author} {\bibfnamefont {J.~F.}\ \bibnamefont
  {Schooley}}, \bibinfo {author} {\bibfnamefont {W.~R.}\ \bibnamefont
  {Hosler}}, \ and\ \bibinfo {author} {\bibfnamefont {M.~L.}\ \bibnamefont
  {Cohen}},\ }\href {\doibase 10.1103/PhysRevLett.12.474} {\bibfield  {journal}
  {\bibinfo  {journal} {Phys. Rev. Lett.}\ }\textbf {\bibinfo {volume} {12}},\
  \bibinfo {pages} {474} (\bibinfo {year} {1964})}\BibitemShut {NoStop}%
\bibitem [{\citenamefont {Pai}\ \emph {et~al.}(2018)\citenamefont {Pai},
  \citenamefont {Lee}, \citenamefont {Lee}, \citenamefont {Annadi},
  \citenamefont {Cheng}, \citenamefont {Lu}, \citenamefont {Tomczyk},
  \citenamefont {Huang}, \citenamefont {Eom}, \citenamefont {Irvin},\ and\
  \citenamefont {Levy}}]{SrTiO31D}%
  \BibitemOpen
  \bibfield  {author} {\bibinfo {author} {\bibfnamefont {Y.-Y.}\ \bibnamefont
  {Pai}}, \bibinfo {author} {\bibfnamefont {H.}~\bibnamefont {Lee}}, \bibinfo
  {author} {\bibfnamefont {J.-W.}\ \bibnamefont {Lee}}, \bibinfo {author}
  {\bibfnamefont {A.}~\bibnamefont {Annadi}}, \bibinfo {author} {\bibfnamefont
  {G.}~\bibnamefont {Cheng}}, \bibinfo {author} {\bibfnamefont
  {S.}~\bibnamefont {Lu}}, \bibinfo {author} {\bibfnamefont {M.}~\bibnamefont
  {Tomczyk}}, \bibinfo {author} {\bibfnamefont {M.}~\bibnamefont {Huang}},
  \bibinfo {author} {\bibfnamefont {C.-B.}\ \bibnamefont {Eom}}, \bibinfo
  {author} {\bibfnamefont {P.}~\bibnamefont {Irvin}}, \ and\ \bibinfo {author}
  {\bibfnamefont {J.}~\bibnamefont {Levy}},\ }\href {\doibase
  10.1103/PhysRevLett.120.147001} {\bibfield  {journal} {\bibinfo  {journal}
  {Phys. Rev. Lett.}\ }\textbf {\bibinfo {volume} {120}},\ \bibinfo {pages}
  {147001} (\bibinfo {year} {2018})}\BibitemShut {NoStop}%
\bibitem [{\citenamefont {Cao}\ \emph {et~al.}(2018)\citenamefont {Cao},
  \citenamefont {Fatemi}, \citenamefont {Fang}, \citenamefont {Watanabe},
  \citenamefont {Taniguchi}, \citenamefont {Kaxiras},\ and\ \citenamefont
  {Jarillo-Herrero}}]{MAGraphene}%
  \BibitemOpen
  \bibfield  {author} {\bibinfo {author} {\bibfnamefont {Y.}~\bibnamefont
  {Cao}}, \bibinfo {author} {\bibfnamefont {V.}~\bibnamefont {Fatemi}},
  \bibinfo {author} {\bibfnamefont {S.}~\bibnamefont {Fang}}, \bibinfo {author}
  {\bibfnamefont {K.}~\bibnamefont {Watanabe}}, \bibinfo {author}
  {\bibfnamefont {T.}~\bibnamefont {Taniguchi}}, \bibinfo {author}
  {\bibfnamefont {E.}~\bibnamefont {Kaxiras}}, \ and\ \bibinfo {author}
  {\bibfnamefont {P.}~\bibnamefont {Jarillo-Herrero}},\ }\href
  {http://dx.doi.org/10.1038/nature26160} {\bibfield  {journal} {\bibinfo
  {journal} {Nature}\ }\textbf {\bibinfo {volume} {556}},\ \bibinfo {pages} {43
  EP } (\bibinfo {year} {2018})}\BibitemShut {NoStop}%
\bibitem [{\citenamefont {Kang}\ \emph {et~al.}(2018)\citenamefont {Kang},
  \citenamefont {Jung}, \citenamefont {Shin}, \citenamefont {Sohn},
  \citenamefont {Ryu}, \citenamefont {Kim}, \citenamefont {Hoesch},\ and\
  \citenamefont {Kim}}]{MoS2}%
  \BibitemOpen
  \bibfield  {author} {\bibinfo {author} {\bibfnamefont {M.}~\bibnamefont
  {Kang}}, \bibinfo {author} {\bibfnamefont {S.~W.}\ \bibnamefont {Jung}},
  \bibinfo {author} {\bibfnamefont {W.~J.}\ \bibnamefont {Shin}}, \bibinfo
  {author} {\bibfnamefont {Y.}~\bibnamefont {Sohn}}, \bibinfo {author}
  {\bibfnamefont {S.~H.}\ \bibnamefont {Ryu}}, \bibinfo {author} {\bibfnamefont
  {T.~K.}\ \bibnamefont {Kim}}, \bibinfo {author} {\bibfnamefont
  {M.}~\bibnamefont {Hoesch}}, \ and\ \bibinfo {author} {\bibfnamefont {K.~S.}\
  \bibnamefont {Kim}},\ }\href {\doibase 10.1038/s41563-018-0092-7} {\bibfield
  {journal} {\bibinfo  {journal} {Nat. Mater.}\ } (\bibinfo {year}
  {2018})}\BibitemShut {NoStop}%
\bibitem [{\citenamefont {Hohenadler}(2016)}]{Mixed_halffilling}%
  \BibitemOpen
  \bibfield  {author} {\bibinfo {author} {\bibfnamefont {M.}~\bibnamefont
  {Hohenadler}},\ }\href {\doibase 10.1103/PhysRevLett.117.206404} {\bibfield
  {journal} {\bibinfo  {journal} {Phys. Rev. Lett.}\ }\textbf {\bibinfo
  {volume} {117}},\ \bibinfo {pages} {206404} (\bibinfo {year}
  {2016})}\BibitemShut {NoStop}%
\bibitem [{\citenamefont {Werman}\ \emph {et~al.}(2017)\citenamefont {Werman},
  \citenamefont {Kivelson},\ and\ \citenamefont {Berg}}]{ErezBerg_nonFermiSSH}%
  \BibitemOpen
  \bibfield  {author} {\bibinfo {author} {\bibfnamefont {Y.}~\bibnamefont
  {Werman}}, \bibinfo {author} {\bibfnamefont {S.~A.}\ \bibnamefont
  {Kivelson}}, \ and\ \bibinfo {author} {\bibfnamefont {E.}~\bibnamefont
  {Berg}},\ }\href {\doibase 10.1038/s41535-017-0009-8} {\bibfield  {journal}
  {\bibinfo  {journal} {npj Quantum Mater.}\ }\textbf {\bibinfo {volume}
  {2}},\ \bibinfo {pages} {7} (\bibinfo {year} {2017})}\BibitemShut {NoStop}%
\end{thebibliography}



%

  \end{document}